\documentclass[12pt]{article}
\usepackage{latexsym}
\usepackage{amssymb}
\usepackage{amsmath}
\usepackage{amsfonts}
\usepackage{eufrak}

\numberwithin{equation}{section}

\newcommand{\be}{\begin{equation}}
\newcommand{\ee}{\end{equation}}

\def\cN{{\cal N}}
\def\bea{\begin{eqnarray}}
\def\eea{\end{eqnarray}}
\def\nn{\nonumber}

\begin{document}

\begin{titlepage}
\phantom{x}

\vspace{1cm}

\begin{center}
{
{\LARGE\bf Two-loop low-energy effective action \\[1mm] in Abelian
supersymmetric\\[3mm]
 Chern-Simons matter models} \vspace{1.5cm}
 \\[0.5cm]
 {\bf
 I.L. Buchbinder $^+$\footnote{joseph@tspu.edu.ru},
 B.S. Merzlikin $^{+\dag}$\footnote{merzlikin@tspu.edu.ru} and
 I.B. Samsonov $^{*}$\footnote{samsonov@mph.phtd.tpu.ru, on leave from
Tomsk Polytechnic University, 634050 Tomsk, Russia.}}
\\[3mm]
 {\it $^+$ Department of Theoretical Physics, Tomsk State Pedagogical
 University,\\ Tomsk 634061, Russia\\[2mm]
 $^\dag$
 Department of Higher Mathematics and Mathematical Physics,\\
\it Tomsk Polytechnic University, 634050 Tomsk, Russia
 \\[2mm]
 $^*$ INFN, Sezione di Padova, via F. Marzolo 8, 35131 Padova, Italy}
 \\[1cm]
}

\begin{abstract}
We compute two-loop low-energy effective actions in Abelian
Chern-Simons matter models with $\cN=2$ and $\cN=3$ supersymmetry up
to four-derivative order. Calculations are performed with a
slowly-varying gauge superfield background. Though the gauge superfield propagator
depends on the gauge fixing parameter, it is shown that the
obtained results are independent of this parameter.
In the massless case the considered models are superconformal. We
demonstrate that
the superconformal symmetry strongly restricts the form of two-loop
quantum corrections to the effective actions such that the obtained
terms have simpler structure than the analogous ones in the
effective action of three-dimensional supersymmetric electrodynamics
(SQED) with vanishing topological mass.
\end{abstract}
\end{center}

\vspace{5mm} Keywords: Effective action, extended supersymmetry,
Chern-Simons model, quantum electrodynamics

\end{titlepage}
\setcounter{footnote}{0}

\section{Introduction}

Three-dimensional gauge field theories have one important
difference from the four-di\-men\-si\-onal ones: they allow for a
gauge invariant topological mass term described by the
Chern-Simons action. In supersymmetric gauge theories, the
Chern-Simons term appears to be crucial in construction of $\cN=8$
and $\cN=6$ superconformal models, known as the BLG
\cite{BL1,BL2,BL3,BL4,G1,G2} and ABJM \cite{ABJM} ones, which are
central objects in the AdS$_4$/CFT$_3$ correspondence. As is
stressed in the recent paper by John Schwarz \cite{Schwarz}, it is
important to study the low-energy effective action in these models
to check the conjecture that it describes the dynamics of probe M2
brane in the AdS$_4$ background.

Leaving the issue of low-energy effective action in ABJM and BLG
models for further studies, in the present paper we consider a
simple problem: what is the dependence of low-energy effective
action in three-dimensional supersymmetric models on
the topological mass ${\mathfrak m}=\frac{kg^2}{2\pi}$, where $g$
is the three-dimensional gauge coupling constant and $k$ is the
Chern-Simons level. There are two special cases, $g\to \infty$
with $k$ finite and $k=0$ with $g$ finite.
The latter corresponds to the gauge theory without the
Chern-Simons term (e.g., SQED or SQCD) while the former case
describes a gauge theory with infinitely large topological mass.
The aim of this paper is to compare the structure of
low-energy effective actions in three-dimensional gauge theories
in these two particular cases.

We address this question
by considering low-energy effective action in Abelian $\cN=2$ supersymmetric
gauge theories with matter. In the
recent paper \cite{BMS1} the two-loop low-energy effective action
in $\cN=2$, $d=3$ SQED (with vanishing topological mass) was computed,
thanks to the background field method in
$\cN=2$, $d=3$ superspace \cite{BPS1,BPS2,BPS-back}. In the
present paper we consider a similar model, but with the
Chern-Simons kinetic term for the gauge superfield rather than the
Maxwell one (i.e., infinitely large topological mass).
We compute two-loop low-energy effective action in
this model up to the four-derivative order and compare it with the
similar terms in the effective action of $\cN=2$, $d=3$ SQED
with vanishing topological mass considered in \cite{BMS1}.
To be more precise, we consider a part of the effective
action which includes only the gauge superfield because these
terms can be naturally compared with the ones studied in
\cite{BMS1}. In general, the effective action involves also
contributions with the chiral matter superfields which are not
considered here. The study of such terms in the effective action is
a separate problem.

The one-loop effective action in gauge superfield sector
(supersymmetric one-loop Euler-Heisenberg effective action) originates from
the loop of matter chiral superfields with external gauge
superfield. It is independent of both couplings $g$ and $k$.
So, we have to consider the two-loop effective
action to study the problem described above. In three-dimensions,
the $\cN=2$ gauge superfield $V$ has not only Grassmann-odd
superfield strengths $W_\alpha$ and $\bar W_\alpha$, but also the
Grassmann-even scalar superfield strength $G$. Up to
four-derivative order, the low-energy effective action for these
superfields has the following structure (see sect.\ \ref{general}
for a more detailed discussion)
\be
\Gamma=\int d^7z \left[
f_1(G) + f_2(G)W^\alpha \bar W^\beta N_{\alpha\beta}
+ f_3(G) W^2 \bar W^2
 \right],
\label{1}
\ee
where $f_i(G)$ are some functions and $N_{\alpha\beta}=D_\alpha
W_\beta$. In the present paper we find two-loop quantum
contributions to the functions $f_i(G)$ and compare them with similar
results in the $\cN=2$, $d=3$ SQED without the Chern-Simons term.


The function $f_1(G)$ in (\ref1) is the leading term in the
low-energy effective action for the gauge superfield. In
components, it is responsible for the $F^2$ terms and its
supersymmetric completions,  with $F_{mn}$ being the Maxwell field
strength. In the $\cN=2$ SQED with the pure Maxwell kinetic term
for the gauge superfield this function has a good geometrical
interpretation: its second derivative defines the moduli space
metric in Coulomb branch \cite{deBoer}. In particular, in
\cite{BMS1} we computed two-loop quantum corrections to the moduli
space in the $\cN=2$ SQED. However, it is known \cite{AHISS} that
the Coulomb branch is absent in three-dimensional gauge theories
with non-trivial Chern-Simons term because the corresponding
equations of motion do not have constant solutions for scalar
fields in the gauge multiplet. In the present paper we show that the function $f_1(G)$
does not receive two-loop quantum corrections in the $\cN=2$
Chern-Simons electrodynamics, but it has non-trivial one-loop
contributions found in \cite{BPS1}. This one-loop contribution to
$f_1(G)$ originates from the loop of chiral superfields with
external gauge superfields and it is independent of whether we
have the super-Maxwell or Chern-Simons propagator for the gauge
superfield.

The functions $f_2(G)$ and $f_3(G)$ in (\ref1) are
responsible for the $F^4$ component term and its supersymmetric
completions. This term is present in the effective
action in both cases, when the gauge superfield is described by
the Maxwell and Chern-Simons terms. Clearly, the form of these
functions $f_1$ and $f_2$ should be different in these two cases.
Indeed, the conventional three-dimensional SQED with the Maxwell
kinetic term for the gauge superfield involves the dimensionful
gauge coupling constant $g$, $[g^2]=1$, such that the model is not
conformal. As a consequence, in the SQED with the Maxwell kinetic
term the functions $f_i(G)$ in (\ref1) are not restricted by the
conformal invariance. On the contrary, the (massless) Chern-Simons
matter theories are superconformal and the form of these functions
is fixed, up to coefficients. We show that the superconformal
invariance requires the vanishing of two-loop quantum corrections
to $f_1$ and $f_2$ in the Chern-Simons matter models while $f_3$
is expressed in terms of superconformal invariants in the $\cN=2$,
$d=3$ superspace costructed in \cite{BPS1}. These results are also
generalized to the Abelian $\cN=2$ Chern-Simons theory with one
chiral matter superfield (in sect.\ \ref{sect3}) and to $\cN=3$
Chern-Simons matter model (in sect.\ \ref{sect4}).

Our general conclusion about the Chern-Simons matter models is
that the structure of low-energy effective action in such theories is
strongly constrained by superconformal invariance. On the contrary, when
the gauge superfield is described by non-conformal supersymmetric
Maxwell term, many new non-conformal terms appear in the low-energy effective
action.

Before starting the main part of the paper, one more comment is in
order. In general, the off-shell effective action is known to be
gauge dependent by construction.
\footnote{The gauge dependence
should not be confused with the gauge invariance of the effective
action. In general, the effective action in gauge theories depends
on gauge fixing conditions which are used for quantization and
correct definition of the path integral. The background field
method is based on special class of gauge fixing conditions (the so
called background field gauges, see e.g.\ \cite{Weinberg} and references therein). The
background field gauges allow one to construct the effective action
which is gauge invariant under the classical gauge
transformations. However, there are infinitely many background
field gauges, for example if $\chi$ is an admissible background
field gauge then $\alpha\chi$ is also admissible background field
gauge with arbitrary real parameter $\alpha$. As a result, the
gauge invariant effective action constructed in framework of
background field method will depend on the parameter $\alpha$.
Therefore it is said that the effective action constructed in
framework of background field method is gauge invariant but gauge
dependent. However, the S-matrix computed on the basis of the
effective action will be completely gauge independent. All these
points are discussed, e.g.,\ in \cite{Vilk}.

%
}
It becomes gauge independent only for background fields satisfying
the effective equations of motion. In the present paper we
consider the low-energy effective action for slowly-varying gauge
superfield background. The conditions determining such a
background coincide with the $\cN=2$ supersymmetric Maxwell
equations, rather than the equations of motion in the Chern-Simons
matter models under considerations. Hence, one can expect that, in
general, the obtained effective action will be gauge dependent. In
particular, the effective action can depend on the gauge-fixing
parameter appearing in the gauge superfield propagator. In our
case, doing two-loop computation we use the gauge superfield
propagator with arbitrary gauge fixing parameter and prove that
the obtained low-energy results are independent of this parameter.
This is a good evidence that the obtained two-loop contributions
to the effective action are, in fact, gauge independent although
they are derived with use of gauge superfield background which
does not solve the classical (and effective) equations of motion.

Throughout this paper we use the $\cN=2$, $d=3$ superspace
notations and conventions introduced in earlier works
\cite{BPS1,BPS2}.

\section{$\cN=2$ Chern-Simons electrodynamics}
\label{sect2}
\subsection{Classical action and propagators}
The classical action of the considered model in $\cN=2$, $d=3$
superspace reads \be S=\frac{k}{2\pi}\int d^7 z\, VG - \int d^7
z\,\left( \bar Q_+ e^{2V} Q_+ + \bar Q_- e^{-2V} Q_- \right) -
\left( m\int d^5z \,Q_+Q_-  + {\it c.c.}\right), \label{action0}
\ee
where $V$ is a gauge superfield with superfield strength $G=\frac i2
\bar D^\alpha D_\alpha V$ and $Q_\pm$ are chiral matter superfields
having opposite charges with respect to the gauge superfield. Here
$m$ is the mass of the chiral superfield and $k$ is the Chern-Simons
level. For $m=0$ this model is superconformal \cite{BPS1}. The
classical action (\ref{action0}) describes $\cN=2$, $d=3$
supersymmetric electrodynamics with Chern-Simons rather than
Maxwell kinetic term for the photon.

To study the effective action in the gauge superfield sector it is
convenient to use the background field method which was developed
for field theories in the $\cN=2$, $d=3$ superspace in
\cite{BPS-back,BP}. We split the gauge superfield $V$ into the
background $V$ and quantum $v$ parts,\footnote{Note that we denote
the background gauge superfield by the same letter $V$ as the
original gauge superfield in the classical action (\ref{action0}).
We hope that it will not lead to any confusions since after the
background-quantum splitting (\ref{splitting}) the original gauge
superfield $V$ never appears.}
\be V \to V+v\,. \label{splitting}
\ee
Upon this splitting the Chern-Simons term in (\ref{action0})
changes as \bea \frac{k}{2\pi}\int d^7z \,VG\to \frac k{2\pi}\int
d^7z \, VG +\frac k\pi \int d^7z\, v\,G +\frac{i k}{4\pi}\int d^7z
\, v D^\alpha \bar D_\alpha v\,, \label{2.9}
\eea
with the background superfields $V$ and $G$ in the r.h.s. The
terms in (\ref{2.9}) which are linear in $v$ are irrelevant
for quantum loop computations. The chiral superfields $Q_\pm$ are
treated as purely quantum and should be integrated out in the
functional integral.

 The operator in
the last term in (\ref{2.9}) is degenerate and requires gauge
fixing, \be f=i\bar D^2v\,,\qquad \bar f=i D^2v\,, \label{gauge} \ee
where $f$ is a fixed chiral superfield. This gauge is usually
accounted by the following gauge fixing term
\cite{Avdeev1,Avdeev2,NG}
\be S_{\rm gf} = \frac{ik\alpha}{8\pi}\int
d^7z\, v(D^2 + \bar D^2) v\,,
\label{Sgf}
\ee
with $\alpha$ being a real parameter. Adding (\ref{Sgf}) to (\ref{action0}) we get the
gauge fixed action for the quantum superfields corresponding to
internal lines of Feynman supergraphs, \bea
S_{\rm quant}&=&S_2 + S_{\rm int}\,,\\
S_2 &=&\int d^7z \left(
\frac{ik}{4\pi}v H v
-\bar {\cal Q}_+ {\cal Q}_+ -\bar{\cal Q}_-{\cal Q}_-
\right)
-\left( m\int d^5z \,{\cal Q}_+{\cal Q}_- + c.c. \right),
\label{2.12}
\\
S_{\rm int}&=&-2\int d^7z\left[
(\bar{\cal Q}_+ {\cal Q}_+ - \bar{\cal Q}_- {\cal Q}_-)v
+(\bar{\cal Q}_+{\cal Q}_+  +\bar{\cal Q}_- {\cal Q}_-)v^2
\right]+O(v^3)\,,
\label{2.13}
\eea
where the operator $H$ reads
\be
H=D^\alpha \bar D_\alpha +\frac\alpha2 (D^2 +\bar D^2)\,.
\label{H}
\ee
In (\ref{2.12}) and (\ref{2.13}) we introduced the notations
${\cal Q}_\pm$ and $\bar{\cal Q}_\pm$ for covariantly
(anti)chiral superfields with respect to the background gauge
superfield,
\be
\bar{\cal Q}_+ = \bar Q_+ e^{2V}\,,\quad
{\cal Q}_+ = Q_+\,,\quad
\bar{\cal Q}_- = \bar Q_- e^{-2V}\,,\quad
{\cal Q}_- = Q_-\,.
\label{cov-Q}
\ee

Let us consider the propagator for the superfield $v$,
\be
2i \langle v(z)v(z')\rangle =G (z,z')\,,
\ee
where the Green's function $G(z,z')$ obeys the equation
\be
\frac{ik}{4\pi}H G(z,z') = -\delta^7(z-z')\,.
\ee
A formal solution to this equation reads
\be
G(z,z')=G_1(z,z')+G_2(z,z')\,,
\label{gauge-prop}
\ee
where
\bea
G_1(z,z')&=&\frac{i\pi}{k}
\frac{D^\alpha \bar D_\alpha}{\square} \delta^7(z-z')
=-\frac\pi k D^\alpha \bar D_\alpha
\int_0^\infty \frac{ds}{(4\pi i
s)^{3/2}}e^{\frac{i\xi^2}{4s}}\zeta^2 \bar\zeta^2
\,,
\label{prop-0}
\\
G_2(z,z')&=&
\frac{i\pi}{2k\alpha} \frac{D^2 +\bar D^2}{\square}
\delta^7(z-z')
=-\frac\pi{2 k\alpha}(D^2 +\bar D^2 )
\int_0^\infty \frac{ds}{(4\pi i
s)^{3/2}}e^{\frac{i\xi^2}{4s}}\zeta^2 \bar\zeta^2
\,.
\label{prop-1}
\eea
Here we applied the standard proper time representation for the
inverse d'Alembertian operator in terms of the components of supersymmetric
interval $\xi^m$ and $\zeta$'s (see the details and references in Appendices A and B).

Note that $G_2(z,z')$ depends on the gauge-fixing parameter
$\alpha$ while $G_1(z,z')$ does not.
We do not fix particular values of this parameter to keep
control on gauge dependence of the effective action.

The action (\ref{2.13}) is responsible for cubic and quartic
interaction vertices while the terms in (\ref{2.12}) give the
propagators for the chiral matter superfields,
\bea
i\langle{\cal Q}_+(z){\cal Q}_-(z')\rangle&=& -m G_+(z,z')\,, \nn \\
i\langle \bar{\cal Q}_+(z) \bar{\cal Q}_-(z')\rangle &=& m G_-(z',z)\,, \nn \\
i\langle {\cal Q}_+(z)\bar {\cal Q}_+(z')\rangle&=& G_{+-}(z,z')=G_{-+}(z',z)\,,\nn\\
i\langle \bar{\cal Q}_-(z) {\cal Q}_-(z')\rangle&=&G_{-+}(z,z')\,.
\label{propagators}
\eea
Properties of Green's functions in the r.h.s.\ of
(\ref{propagators}) were studied in \cite{BMS1,BPS1}. Explicit
expressions for them are given in Appendix \ref{AppB}.

\subsection{General structure of effective action}
\label{general}
Our aim is to study the low-energy effective action in the model
(\ref{action0}) in the gauge superfield sector.
It can be written
as
\be
\Gamma=S_{\rm cl}+\bar\Gamma\,,
\ee
where $S_{\rm cl}= \frac{k}{2\pi}\int d^7 z\, VG$ is the classical
Chern-Simons term and $\bar\Gamma$ takes into account quantum
corrections to the effective action. In what follows we will consider only $\bar\Gamma$
omitting `bar' for brevity.

In general, $\Gamma$
is a functional of superfield strengths $G$, $W_\alpha$, $\bar
W_\alpha$ and their derivatives, $N_{\alpha\beta}= D_\alpha
W_\beta$, $\bar N_{\alpha\beta}=\bar D_\alpha \bar W_\beta$,
\be
\Gamma = \int d^7z\,{\cal L}(G,W_\alpha,\bar W_\alpha,N_{\alpha\beta},\bar N_{\alpha\beta},
\ldots)\,,
\label{eff0}
\ee
where dots stand for higher-order derivatives of the superfield
strengths.
It is very difficult to find the effective action
(\ref{eff0}) taking into account all derivatives of the fields.
Therefore, to simplify the problem, we restrict ourself to the
terms with no more than four space-time derivatives of component
fields. A typical bosonic representative in components is
$f(\phi)(F^{mn}F_{mn})^2$, where $F_{mn}$ is the Maxwell field
strength and $f(\phi)$ is some function of the scalar field $\phi$
which is part of the $\cN=2$, $d=3$ gauge multiplet. It is clear
that to find this term in the effective action it is sufficient to
consider constant fields $F_{mn}$ and $\phi$. In terms of
superfields, such a background corresponds to the following
constraints on the superfield strengths:
\begin{itemize}
\item[(i)] Supersymmetric Maxwell equations,
\be
D^\alpha W_\alpha = 0\,,\qquad \bar D^\alpha \bar W_\alpha =0\,;
\label{c1}
\ee
\item[(ii)] Superfield strengths are constant with respect to the space-time
coordinates,
\be
\partial_m G = \partial_m W_\alpha = \partial_m \bar
W_\alpha=0\,.
\label{c2}
\ee
\end{itemize}
We emphasize that though the eqs.\ (\ref{c1})  are not the equations of
motion in the theory under consideration, they, together with
eqs.\ (\ref{c2}), single out the slowly varying gauge superfield background.
In components, such a background contains constant scalar $\phi$,
spinor $\lambda_\alpha$, $\bar\lambda_\alpha$ and Maxwell $F_{mn}$ fields
while the auxiliary field $D$ vanishes owing to (\ref{c1}).
For the gauge superfield background constrained by (\ref{c1}) and (\ref{c2}) we
can use the exact expressions for the chiral superfield propagators (\ref{B7}),
(\ref{B8}) and (\ref{B9}) which were derived in \cite{BMS1}.

Note that
the superfields $N_{\alpha\beta}$ and $\bar N_{\alpha\beta}$
and not independent subject to the constraints (\ref{c1}) and
(\ref{c2}),
\be
N_{\alpha\beta}=-\bar N_{\alpha\beta}\,.
\ee
Hence, we keep only $N_{\alpha\beta}$ and discard $\bar N_{\alpha\beta}$ in what
follows assuming that the latter is expressed from the former.

Under the constraints (\ref{c1}) and (\ref{c2}) the effective
action (\ref{eff0}) in components contains Maxwell field strength
in arbitrary power and, so, involves arbitrary number of
space-time derivatives. The superfield action which contains the terms
with no more than four derivatives is given by
\be
\Gamma=\int d^7z \left[
f_1(G) + f_2(G)W^\alpha \bar W^\beta N_{\alpha\beta}
+ f_3(G) W^2 \bar W^2
 \right],
\label{eff1}
\ee
with some functions $f_i(G)$, $i=1,2,3$. Indeed, the full
superspace measure $d^7z$ involves the Grassmann-odd coordinate
part $d^2\theta d^2\bar\theta\propto D^2 \bar D^2$. Thus, it
counts as two space-time derivatives. Next, $W^2 \bar W^2$ also
contain effectively four $D$'s (which count as two $\partial_m$)
because of $W_\alpha = \bar D_\alpha G$ and $\bar W_\alpha = D_\alpha
G$. Hence, the first term in the r.h.s.\ of (\ref{eff1}) is a
two-derivative piece while the other terms are four-derivative
ones.

In principle, one could
include in (\ref{eff1}) also the term of the form
$\int d^7z \, f(G)W^\alpha\bar W_\alpha$, but it vanishes for the
gauge superfield background subject to (\ref{c1}),
\be
\int d^7z \, f(G) W^\alpha \bar W_\alpha
=-\frac12 \int d^5z \, (\bar D^\alpha f(G))(\bar D_\alpha \bar
W^\beta )W_\beta
=-\frac 12\int d^5z \, W^\alpha N_{\alpha}^\beta W_\beta
f'(G)=0\,.
\label{f-zero}
\ee
Here we passed from the full superspace to the chiral measure and
used the fact that $N^\alpha_\beta$ is traceless,
$N^\alpha_\alpha=0$, subject to (\ref{c1}).

Let us discuss the component structure of the effective action
(\ref{eff1}) in the bosonic sector. For this purpose it is
sufficient to consider the gauge superfield $V$ of the special
form:
\be
\hat V= i\theta^\alpha \bar\theta_\alpha \phi +\theta^\alpha \bar\theta^\beta
\gamma^m_{\alpha\beta} A_m\,,
\ee
where $\phi$ is a constant scalar and $A_m$ is a gauge vector
field with constant Maxwell field strength, $F_{mn}=\partial_m A_n - \partial_n
A_m$. The superfield strengths constructed with the use of this
gauge superfield have the following component structure
\bea
\hat G &=& -\phi -\frac12\varepsilon^{mnp}(\gamma_p)^{\alpha\beta}
\theta_\alpha \bar\theta_\beta F_{mn}\,,\\
\hat W_\alpha &=& \frac12 \varepsilon^{mnp}(\gamma_p)_\alpha^\beta
\theta_\beta F_{mn}\,,\qquad
\hat{\bar W}_\alpha = \frac12\varepsilon^{mnp}(\gamma_p)_\alpha^\beta
\bar\theta_\beta F_{mn}\,.
\eea
With these superfields, we find that the effective action
(\ref{eff1}) contains the following terms in its component field
decomposition
\be
\Gamma = \frac18 \int d^3x \left\{
f''_1(-\phi)F^{mn}F_{mn}
+[2f_3(-\phi)-f'_2(-\phi)](F^{mn}F_{mn})^2
 \right\}+\ldots,
\label{eff2}
\ee
where dots stand for other components which are related with the
given ones by $\cN=2$ supersymmetry. The equation (\ref{eff2})
shows that the first term in r.h.s.\ of (\ref{eff1}) is responsible for the
$F^2$ term while the terms with the functions $f_2$ and $f_3$
result in the $F^4$ term.

In the present paper we will perturbatively compute the functions
$f_i$ in (\ref{eff1}) in the two-loop approximation, \be f_i(G) =
f_i^{(1)}(G) + f_i^{(2)} (G)\,, \ee where $f_i^{(1)}(G)$ and
$f_i^{(2)} (G)$ correspond to one- and two-loop contributions,
respectively. Note that at the one-loop order the effective action
(\ref{eff1}) receives contributions from the loop of (anti)chiral
matter fields only. These contributions were calculated in
\cite{BPS1}\footnote{The function (\ref{f11}) was introduced for
three-dimensional gauge theories in \cite{HKLR} in the study of
non-linear sigma-models with extended supersymmetry. In four
dimensions, analogous function corresponds to the Lagrangian of
improved tensor multiplet (see e.g. \cite{book}).}:
\bea \label{f11}
f_1^{(1)} &=& \frac1{2\pi} (G\ln (G+\sqrt{G^2 + m^2})-
\sqrt{G^2+m^2})\,,\\
f_2^{(1)} &=&0\,,\\
f_3^{(1)} &=& \frac1{128\pi} \frac1{(G^2 +m^2)^{5/2}}\,.
\eea
Our aim now is to find the functions $f^{(2)}_i$ which take
into account two-loop quantum contributions to the effective
action (\ref{eff1}).

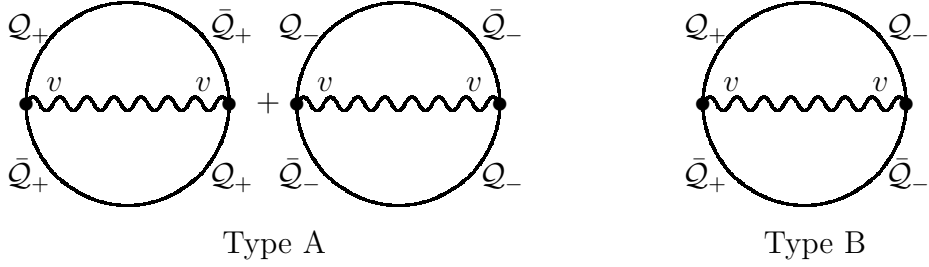
\begin{figure}[t]
\begin{center}
\setlength{\unitlength}{0.9mm}
\begin{picture}(150,50)
\thicklines
\qbezier(35,25)(35,30.74)(30.6,35.6)
\qbezier(30.6,35.6)(25.75,40)(20,40)
\qbezier(20,40)(14.26,40)(9.39,35.6)
\qbezier(9.39,35.6)(5,30.74)(5,25)
\qbezier(5,25)(5,19.26)(9.39,14.39)
\qbezier(9.39,14.39)(14.26,10)(20,10)
\qbezier(20,10)(25.74,10)(30.6,14.39)
\qbezier(30.6,14.39)(35,19.26)(35,25)
\qbezier(75,25)(75,30.74)(70.6,35.6)
\qbezier(70.6,35.6)(65.75,40)(60,40)
\qbezier(60,40)(54.26,40)(49.39,35.6)
\qbezier(49.39,35.6)(45,30.74)(45,25)
\qbezier(45,25)(45,19.26)(49.39,14.39)
\qbezier(49.39,14.39)(54.26,10)(60,10)
\qbezier(60,10)(65.74,10)(70.6,14.39)
\qbezier(70.6,14.39)(75,19.26)(75,25)
\qbezier(135,25)(135,30.74)(130.6,35.6)
\qbezier(130.6,35.6)(125.75,40)(120,40)
\qbezier(120,40)(114.26,40)(109.39,35.6)
\qbezier(109.39,35.6)(105,30.74)(105,25)
\qbezier(105,25)(105,19.26)(109.39,14.39)
\qbezier(109.39,14.39)(114.26,10)(120,10)
\qbezier(120,10)(125.74,10)(130.6,14.39)
\qbezier(130.6,14.39)(135,19.26)(135,25)
\put(5,25){\circle*{2}}
\put(35,25){\circle*{2}}
\put(45,25){\circle*{2}}
\put(75,25){\circle*{2}}
\put(105,25){\circle*{2}}
\put(135,25){\circle*{2}}
\put(39,24){$+$}
\put(2,35){${\cal Q}_+$}
\put(32,35){$\bar{\cal Q}_+$}
\put(42,35){${\cal Q}_-$}
\put(72,35){$\bar{\cal Q}_-$}
\put(102,35){${\cal Q}_+$}
\put(132,35){${\cal Q}_-$}
\put(2,13){$\bar{\cal Q}_+$}
\put(32,13){${\cal Q}_+$}
\put(42,13){$\bar{\cal Q}_-$}
\put(72,13){${\cal Q}_-$}
\put(102,13){$\bar{\cal Q}_+$}
\put(132,13){$\bar{\cal Q}_-$}
\put(8,27){$v$}
\put(30,27){$v$}
\put(48,27){$v$}
\put(70,27){$v$}
\put(108,27){$v$}
\put(130,27){$v$}
\qbezier(5,25)(6,27)(7,25)
\qbezier(7,25)(8,23)(9,25)
\qbezier(9,25)(10,27)(11,25)
\qbezier(11,25)(12,23)(13,25)
\qbezier(13,25)(14,27)(15,25)
\qbezier(15,25)(16,23)(17,25)
\qbezier(17,25)(18,27)(19,25)
\qbezier(19,25)(20,23)(21,25)
\qbezier(21,25)(22,27)(23,25)
\qbezier(23,25)(24,23)(25,25)
\qbezier(25,25)(26,27)(27,25)
\qbezier(27,25)(28,23)(29,25)
\qbezier(29,25)(30,27)(31,25)
\qbezier(31,25)(32,23)(33,25)
\qbezier(33,25)(34,27)(35,25)
\qbezier(45,25)(46,27)(47,25)
\qbezier(47,25)(48,23)(49,25)
\qbezier(49,25)(50,27)(51,25)
\qbezier(51,25)(52,23)(53,25)
\qbezier(53,25)(54,27)(55,25)
\qbezier(55,25)(56,23)(57,25)
\qbezier(57,25)(58,27)(59,25)
\qbezier(59,25)(60,23)(61,25)
\qbezier(61,25)(62,27)(63,25)
\qbezier(63,25)(64,23)(65,25)
\qbezier(65,25)(66,27)(67,25)
\qbezier(67,25)(68,23)(69,25)
\qbezier(69,25)(70,27)(71,25)
\qbezier(71,25)(72,23)(73,25)
\qbezier(73,25)(74,27)(75,25)
\qbezier(105,25)(106,27)(107,25)
\qbezier(107,25)(108,23)(109,25)
\qbezier(109,25)(110,27)(111,25)
\qbezier(111,25)(112,23)(113,25)
\qbezier(113,25)(114,27)(115,25)
\qbezier(115,25)(116,23)(117,25)
\qbezier(117,25)(118,27)(119,25)
\qbezier(119,25)(120,23)(121,25)
\qbezier(121,25)(122,27)(123,25)
\qbezier(123,25)(124,23)(125,25)
\qbezier(125,25)(126,27)(127,25)
\qbezier(127,25)(128,23)(129,25)
\qbezier(129,25)(130,27)(131,25)
\qbezier(131,25)(132,23)(133,25)
\qbezier(133,25)(134,27)(135,25)
\put(34,3){Type A}
\put(114,3){Type B}
  \end{picture}
\end{center}
\caption[b]{Two-loop supergraphs in $\cN=2$ supersymmetric electrodynamics.}
\label{fig1}
\end{figure}

The two-loop effective action is given by the following formal
expression
\bea
\label{Gam}
\Gamma^{(2)}&=&\Gamma_{\rm A}+\Gamma_{\rm B}\,,\\
\Gamma_{\rm A}&=& -2\int d^7z\, d^7z'
G_{+-}(z,z')G_{-+}(z,z')G(z,z')\,,
\label{GA}\\
\Gamma_{\rm B}&=& -2m^2\int d^7z\, d^7z'
 G_+ (z,z')G_-(z,z')G(z,z')\,.
\label{GB}
\eea
The two terms $\Gamma_{\rm A}$ and $\Gamma_{\rm B}$ are represented by
corresponding Feynman graphs in fig.\ \ref{fig1}.

Note that, in general, in the two-loop effective action the
diagrams of topology ``eight'' are also present. Such diagrams
involve either $G_{+-}$ or $G_+$ propagator and the gauge
superfield propagator (\ref{gauge-prop}) which should be
considered at coincident superspace points. However, at coincident
points the gauge superfield propagator (\ref{gauge-prop}) vanishes,
$G(z,z)=0$.
Hence, there are no contributions to the effective action from
the graphs of topology ``eight''.

\subsection{Independence of two-loop effective action of the
gauge-fixing parameter}
The part of the gauge superfield propagator which depends on the
gauge fixing parameter $\alpha$ is given by (\ref{prop-1}). In this
section we will demonstrate that the two-loop contributions to the
low-energy effective action of the form (\ref{eff1}) are independent of this
parameter. To prove this, we check the vanishing of contributions to the
two-loop effective actions (\ref{GA}) and (\ref{GB}) which correspond to the
propagator (\ref{prop-1}).

Consider first the part of the effective action (\ref{GA}).
The propagator (\ref{prop-1}) contains the operator $D^2+\bar D^2$
acting on the full superspace delta-function. With the use of
integration by parts, the operator $\bar D^2$ hits the Green's function
$G_{-+}(z,z')$ (similarly, $D^2$ hits $G_{+-}(z,z')$).
According to (\ref{A5}), one gets two terms:
\be
\frac14\bar\nabla^2 G_{-+}(z,z') =
-\delta_+(z,z')-m^2 G_+(z,z')\,.
\label{3.5}
\ee
The delta-function in (\ref{3.5}) gives vanishing contribution to
(\ref{GA}) since the expression (\ref{prop-1}) already contains the
Grassmann delta-function $\zeta^2\bar\zeta^2 =
\delta^2(\theta-\theta')\delta^2(\bar\theta-\bar\theta')$.

Consider the contributions to $\Gamma_{\rm A}$ from the last term in
(\ref{3.5}). With the use of the heat kernel representations of the propagators
(\ref{B7})--(\ref{B8}), the part of the effective action
corresponding to the last term in (\ref{3.5}) reads
\bea
&&\int d^7z d^7 z' \int_0^\infty\frac{ds\,dt\,du}{(4\pi iu)^{3/2}}
\zeta^2 \bar\zeta^2 e^{\frac{i\xi^2}{4u}}
e^{i(s+t)m^2}
K_+(z,z'|s)K_{+-}(z,z'|t)
\nn\\
&=&\int d^7z d^3\xi
\int_0^\infty\frac{ds\,dt\,du}{(4\pi iu)^{3/2}}
e^{\frac{i\xi^2}{4u}}e^{i(s+t)m^2}
K_+(z,z'|s)K_{+-}(z,z'|t)\Big|\,.
\label{3.6_}
\eea
Here we integrated over one set of Grassmann variables using the
delta-function. The symbol $\big|$ in the second line of
(\ref{3.6_}) means that this expression is considered at
coincident Grassmann coordinates,
\be
\Big| \equiv \Big |_{\theta=\theta'\,,
\
\bar\theta=\bar\theta'\,.}
\label{||}
\ee
Note that the bosonic coordinates $x_m$ and $x_m'$ remain
different under this projection. We will employ
the notation (\ref{||}) throughout the present paper.

 It is important to note that the heat kernel
$K_+$ at coincident superspace points contains $W^2$, see
(\ref{K+limit}). Hence, the result of calculation of the expression
(\ref{3.6_}) can always be represented in the form
\be
\int d^7z\, W^2 {\cal F}(G)\,,
\label{3.7_}
\ee
with some function ${\cal F}(G)$. One can easily see that the
quantity (\ref{3.7_}) vanishes for the on-shell gauge superfield (\ref{c1}).
Indeed, passing to the chiral subspace one gets
\be
\int d^7z\, W^2 {\cal F}(G)
=-\frac14 \int d^5z \, W^2\bar D^2 {\cal F}(G)
=-\frac14\int d^5z \, W^2 W^2 {\cal F}''(G)
\equiv0\,.
\ee
This expression vanishes as it contains too many Grassmann-odd superfields
$W_\alpha$.

Consider now the contributions to the effective action $\Gamma_{\rm B}$
from the propagator (\ref{prop-1}). Similarly as for $\Gamma_{\rm
A}$, after integration by parts, the operator $\bar D^2$ hits $K_-$ and
produces $K_{+-}$ because of the identity
\be
K_{+-}(z,z'|s)= \frac14\bar\nabla^2 K_-(z,z'|s)\,.
\ee
Hence, the part of the effective action $\Gamma_{\rm B}$ gets the
same form (\ref{3.6_}) and, thus, vanishes.

The present analysis was done for the operator $\bar D^2$
in (\ref{prop-1}). The operator $D^2$ can be considered in a
similar way with the same conclusion. Thus, we proved that the two-loop contributions to
the effective action (\ref{eff1}) with the propagator
(\ref{prop-1}) vanish. In other words, the considered low-energy
effective action is independent of the gauge-fixing parameter
$\alpha$. In the following sections we will compute
non-trivial contributions to the two-loop effective actions
(\ref{GA}) and (\ref{GB}) coming from the gauge superfield
propagator $G_1$ given by (\ref{prop-0}).

\subsection{Two-loop graph A}
\label{sectA}
Consider the part of the effective action (\ref{GA}) and represent
all the Green's functions in terms of the corresponding heat
kernels,
\be
\Gamma_{\rm A}=-\frac{2\pi}k\int d^7z\, d^3\xi
\int_0^\infty \frac{ds\,dt\, du}{(4\pi i u)^{3/2}}e^{\frac{i\xi^2}{4u}}
e^{i(s+t)m^2}
\nabla^\alpha K_{+-}(z,z'|s)\bar \nabla_\alpha K_{-+}(z,z'|t)
\Big|\,.
\label{3.10}
\ee
Here we integrated by pats the derivatives $D^\alpha \bar
D_\alpha$ which come from the gauge superfield propagator (\ref{prop-0}). To find the
effective action we need to compute the derivatives of the heat
kernels,
$\nabla_\alpha K_{+-}(z,z'|s)$ and $\bar \nabla_\alpha K_{-+}(z,z'|t)$.
In general, this problem is very
hard since the heat kernels themselves have very complicated form
(\ref{K_-+2_}) and (\ref{K_-+3_}). However, we will take into account the
following simplifications:
\begin{itemize}
\item Upon computing the derivative of the heat kernels we
omit the terms which vanish in the limit $\theta=\theta'$, $\bar\theta=\bar\theta'$.
\item Since we are interested in the low-energy effective action
of the form (\ref{eff1}), it is sufficient to consider only the terms which
depend on superfield strengths $G$, $W_\alpha$, $\bar W_\alpha$,
but which contain $N_{\alpha\beta}$ at most in the first
power. Terms with higher orders of $N_{\alpha\beta}$ should be
systematically neglected.
\end{itemize}
For instance, the formulas (\ref{id's}) up to the first order in
$N_{\alpha\beta}$ read
\bea
W^\alpha(s)&\approx&W^\alpha - sN^\alpha_\beta W^\beta\,,\qquad
\bar W^\alpha(s)\approx \bar W^\alpha - sN^\alpha_\beta \bar
W^\beta\,,\\
\zeta^\alpha(s)&\approx&\zeta^\alpha
-sW^\alpha+\frac12 s^2 N^\alpha_\beta W^\beta\,,\\
\bar \zeta^\alpha(s) &\approx &\bar\zeta^\alpha -s\bar W^\alpha
+\frac12 s^2 N^\alpha_\beta \bar W^\beta\,,\\
\xi^m(s)&\approx&\xi^m-i(\gamma^m)^{\alpha\beta}
[s(W_\alpha\bar\zeta_\beta + \bar W_\alpha \zeta_\beta)
\nn\\&&
-\frac{s^2}2 N_{\alpha\gamma}(W^\gamma\bar\zeta_\beta + \bar W^\gamma \zeta_\beta)
+\frac{s^3}6 N_{\alpha\beta}W\bar W]\,.
\label{3.14}
\eea
Here and further the symbol ``$\approx$'' means that the
expressions are considered in the corresponding approximation up to the
first order in $N_{\alpha\beta}$ and all terms of order $O(N^2)$ are omitted.

To compute the expression (\ref{3.10}) we have to find $\nabla_\alpha
K_{+-}(z,z'|s)|$ and
$\bar\nabla_\alpha K_{-+}(z,z'|t)|$. Using (\ref{K_-+2_})
these quantities can be recast as
\bea
\nabla_\alpha K_{+-}(z,z'|s)| &=& M_\alpha(s) \cdot
K_{+-}(z,z'|s)|\,,
\nn\\
\bar\nabla_\alpha K_{-+}(z,z'|t)|
&=&\tilde M_\alpha (t)\cdot K_{-+}(z,z'|t)|\,,
\label{3.15}
\eea
where
\bea
M_\alpha(s) &=& \Big[
2i s G\bar W_\alpha +\frac i2 (F\coth(sF))_{mn}\rho^m(s)\nabla_\alpha \rho^n(s) +
\nabla_\alpha R(z,z') +\nabla_\alpha I(z,z')
\nn\\&&
+\int_0^s d\tau \nabla_\alpha(R'(\tau)+\Sigma(\tau))
\Big]\Big|\,,
\label{3.16_}\\
\tilde M_\alpha(t) &=& \Big[
2i t G W_\alpha +\frac i2 (F\coth(tF))_{mn}\tilde\rho^m(t)\bar\nabla_\alpha \tilde\rho^n(t) +
\bar\nabla_\alpha \tilde R(z,z') +\bar\nabla_\alpha I(z,z')
\nn\\&&
+\int_0^t d\tau \bar\nabla_\alpha(\tilde R'(\tau)+\Sigma(\tau))
\Big]\Big|.
\label{3.16}
\eea
Here $\rho^m$ and $\tilde\rho^m$ are versions of the bosonic interval
with specific chirality properties (\ref{chirho}). The two-point
quantities $R(z,z')$, $\tilde R(z,z')$ and $\Sigma(z,z')$ are
written down explicitly in (\ref{R}), (\ref{tildeR}) and
(\ref{Sigma}), respectively. Basic properties of the parallel transport
propagator $I(z,z')$ are summarized in Appendix \ref{appA}.

Using the equations (\ref{3.14}), (\ref{DI1}), (\ref{R}) and
(\ref{RS}) we compute derivatives of various objects in (\ref{3.16_})
and (\ref{3.16}),
\bea
\nabla_\alpha  \rho^m(s)|&\approx& is^2 \gamma^m_{\beta\gamma}
 N_\alpha^\beta \bar W^\gamma\,,\label{e1}\\
\nabla_\alpha R(z,z')|&\approx&
-\frac12\xi_{\alpha\beta}\bar W^\beta\,,\label{e2}\\
\nabla_\alpha I(z,z')|&\approx&
\frac12\xi_{\alpha\beta}\bar W^\beta I(z,z')\,,\label{e3}\\
\int_0^s d\tau \,\nabla_\alpha(R'(\tau)+\Sigma(\tau))|
&\approx& i s^2 G N_{\alpha\beta }\bar W^\beta
+2i s^2 \bar W^2 W_\alpha\,.\label{e4}
\eea
One can easily find similar expressions involving the derivative
$\bar\nabla_\alpha$ in the l.h.s.  Substituting (\ref{e1})--(\ref{e4}) into (\ref{3.16_}) we get
\bea
M_\alpha(s)&\approx& 2is G\bar W_\alpha
+is^2GN_{\alpha\beta}\bar W^\beta
+2i s^2 \bar W^2 W_\alpha
-\frac s2 \xi_m\gamma^m_{\beta\gamma} N_\alpha^\beta \bar
W^\gamma
-\frac{3i}4 s^3 \bar W^2 N_{\alpha\beta}W^\beta\,,
\nn\\
\tilde M_\alpha(t)&\approx& 2it G W_\alpha
 +it^2 G N_{\alpha\beta} W^\beta
 -2it^2 W^2 \bar W_\alpha
 +\frac t2 \xi_m \gamma^m_{\beta\gamma} N_\alpha^\gamma W^\beta
 +\frac{3i}4 t^3 W^2 N_{\alpha\beta}\bar W^\beta\,.
 \nn\\
 \label{MM}
\eea

The equations (\ref{3.15}) include also the heat kernels
$K_{+-}$ and $K_{-+}$
at coincident Grassmann points (\ref{kerlim}). We have to expand
(\ref{kerlim}) up to the first order in $N_{\alpha\beta}$. In
particular, the functions (\ref{fun}) in this approximation are
\bea
f_\alpha{}^\beta(s)&\approx& - s^2 \delta_\alpha^\beta + \frac13
s^3 N_\alpha^\beta\,,\\
f(s)&\approx& -\frac 7{12}s^3\,,\\
f_{\alpha\beta}^m(s)&\approx& -\frac{s}{2}\gamma^m_{\alpha\beta}
+\frac1{12}s^2 \varepsilon_{\alpha\beta} (\gamma^m_{\rho\sigma}N^{\rho\sigma})
+\frac34 s^2(\gamma^m_{\beta\gamma}N_\alpha^\gamma +
\gamma^m_{\alpha\gamma}N_\beta^\gamma)\,.
\eea
Substituting these functions into (\ref{kerlim}) we find
\be
K_{+-}(z,z'|s)\big|\approx
-\frac1{(4i\pi s)^{3/2}}e^{\frac{i}{4s}\xi^2 + isG^2}
e^{X(\xi^m,s)}\,,
\label{Kapprox}
\ee
where
\bea
X(\xi^m,s)&=&is^2 G W^\alpha\bar W_\alpha
 -\frac i3 s^3 G W^\alpha N_\alpha^\beta \bar W_\beta
 -\frac s2 \xi_m \gamma^m_{\alpha\beta} W^\alpha \bar W^\beta
 \nn\\&&
 +\frac1{12} s^2 \xi_m(\gamma^m N) W^\alpha\bar W_\alpha
 +\frac32 s^2 \xi_m \gamma^m_{\gamma(\alpha} N_{\beta)}^\gamma
  W^\alpha \bar W^\beta
 -\frac{7i}{24}s^3 W^2 \bar W^2\,.
 \label{X}
\eea

With the use of (\ref{3.15}) and (\ref{Kapprox})
the part of the effective action (\ref{3.10}) can be recast as
\bea
\Gamma_{\rm A}&=&-\frac{2\pi}{k(4\pi i)^{9/2}}
\int d^7z d^3\xi
\int_0^\infty \frac{ds\,dt\,du}{(stu)^{3/2}}e^{\frac{i\xi^2}{4}(\frac1s+\frac1t+\frac1u)}
e^{i(s+t)(G^2+m^2)}
\nn\\&&
\times M^\alpha(s)\tilde M_\alpha(t) e^{X(\xi^m,s)+ X(-\xi^m,t)}\,.
\label{3.29}
\eea
The expression in the second line in (\ref{3.29}) should be
expanded in a series up to the first order in $N_{\alpha\beta}$,
\bea
\label{3.30}
M^\alpha(s)\tilde M_\alpha(t) e^{X(\xi^m,s)+ X(-\xi^m,t)}&
\approx&-4 st G^2 W^\alpha \bar W_\alpha
+2st G^2 (s-t)\bar W^\alpha W^\beta N_{\alpha\beta}
\\&&
+4st(t-s) GW^2 \bar W^2
+2ist(s^2+t^2)G^3 W^2 \bar W^2
\nn\\
&&+ist G\xi_m \gamma^m_{\rho\sigma}N_\alpha^\sigma
(\bar W^\alpha W^\rho +\bar W^\rho W^\alpha)
\nn\\&&
+st[\frac i2(s+t)+
\frac{G^2}{12}(s-t)(5s-t)]
 \xi_m (\gamma^m N) W^2 \bar W^2\,.\nn
\eea
Here we used explicit forms of the quantities $M_\alpha(s)$ and
$X(\xi^m,s)$ given in (\ref{MM}) and (\ref{X}), respectively.
The terms in the last two lines in (\ref{3.30}) contain bosonic
interval $\xi^m$ in the first power. They do not
contribute to the effective action because of the identity
\be
\int
d^3\xi\,\xi_m\,e^{\frac{i\xi^2}{4}(\frac1s+\frac1t+\frac1u)}=0\,.
\label{int-zero}
\ee
For the terms in the first two lines in (\ref{3.30}) the
integration over $d^3\xi$ is simply Gaussian,
\be
\int d^3\xi\, e^{\frac i4
a\xi^2}=-\left(\frac{4i\pi}{a}\right)^{\frac32}\,,\qquad
a= \frac1s+\frac1t+\frac1u\,.
\label{d3rho}
\ee
Hence, after integration over $du$, the effective action
(\ref{3.29}) can be recast as
\bea
\Gamma_{\rm A}&=&\frac i{16\pi^2 k}\int d^7z
\int_0^{\infty}ds\,dt\frac{\sqrt{st}}{s+t}e^{i(s+t)(G^2+m^2)}
\big[-4G^2 \bar W^\alpha W_\alpha
\nn\\&&
+2(s-t)G^2\bar W^\alpha W^\beta N_{\alpha\beta}
-4(s-t)G W^2\bar W^2
+2i(s^2+t^2)G^3 W^2\bar W^2
\big].
\label{3.33}
\eea

The expression (\ref{3.33}) contains the term with $W^\alpha \bar
W_\alpha$. This term vanishes on shell because of (\ref{f-zero}).
There are also two terms in (\ref{3.33}) containing $(s-t)$. These
terms are also vanishing since they are odd under the change of
integration variables $s\leftrightarrow t$. So, only the last term
in (\ref{3.33}) remains non-trivial for the considered gauge
superfield background. Performing the integration over $s$ and $t$
in this term we get the final result for the effective
action $\Gamma_{\rm A}$:
\be
\Gamma_{\rm A}= -\frac{15}{256\pi k}\int d^7z \frac{G^3 W^2\bar
W^2}{(G^2+m^2)^4}\,.
\label{3.34}
\ee

\subsection{Two-loop graph B}
\label{SectB}
Consider the part of the effective action (\ref{GB}) with the
gauge superfield propagator (\ref{prop-0}),
\be
\Gamma_{\rm B}=-\frac{2\pi m^2}{k} \int d^7z\, d^3\xi
\int_0^\infty \frac{ds\,dt\,du}{(4\pi i u)^{3/2}}
e^{\frac{i\xi^2}{4u}}e^{i(s+t)m^2}
\nabla^\alpha K_+(z,z'|s)\bar\nabla_\alpha K_-(z,z'|t)
\Big|\,.
\label{3.35}
\ee
Here we integrated by parts the operator $D^\alpha\bar D_\alpha$
and integrated out one set of Grassmann variables using the
delta-function.
For computing this part of the effective action we need to find
the derivatives of the heat kernels (\ref{K+fin}) and (\ref{K-fin})
at coincident Grassmann points,
\be
\nabla_\alpha K_+(z,z'|s)\big|
=
\frac{1}{(4\pi is)^{3/2}}P_\alpha(s)
e^{Y(s)} e^{isG^2}e^{\frac{i\xi^2}{4s}} I(z,z')\big|\,,
\label{DK+}\\
\ee
where
\bea
Y(s)&=&\frac{i}{4}(F\coth(sF))_{mn}\xi^m(s)\xi^n(s)
-\frac{i\xi^2}{4s}\nn\\&&
- \frac12\bar\zeta^\beta(s) \xi_{\beta\gamma}(s)W^\gamma(s)
+\int_0^s dt\, \Sigma(z,z'|t)\,,\\
P_\alpha(s)&=&\nabla_\alpha \zeta^2(s)+ \zeta^2(s)\nabla_\alpha Y(s)
\,.
\label{3.36}
\eea
It is sufficient to compute the derivatives of all objects in
(\ref{3.36}) up to the first order in $N_{\alpha\beta}$,
\bea
\nabla_\alpha \xi^m(s)\big|&\approx&i s\gamma^m_{\alpha\beta}
\bar W^\beta -\frac{i s^2}2\gamma^m_{\alpha\beta}N^\beta_\gamma
\bar W^\gamma\,,\\
\nabla_\alpha \zeta^2(s)\big|&\approx&
-2s W_\alpha - s^2 N_{\alpha\beta}W^\beta\,,\\
-\frac12\nabla_\alpha(\bar\zeta^\beta(s)\xi_{\beta\gamma}(s)W^\gamma(s))\big|
&\approx&-\frac s2 \bar W^\beta \xi_m \gamma^m_{\beta\gamma}
 N_\alpha^\gamma
+\frac{3i}4(s^2 \bar W^2 W_\alpha
 -s^3 \bar W^2 N_{\alpha\beta} W^\beta)\,,\\
\int_0^s dt \nabla_\alpha\Sigma(z,z'|t)\big|
&\approx&-is G \bar W_\alpha
+\frac{is^2}{2}G N_{\alpha\beta}\bar W^\beta
\nn\\&&
-\frac{is^3}{6}W^2N_{\alpha\beta}W^\beta
-\frac s{12}\xi_m(\gamma^m N)\bar W_\alpha\,.
\eea
Substituting these formulas to (\ref{3.36})
 and expanding up to
the first order in $N_{\alpha\beta}$ we get
\bea
P_\alpha(s)e^{Y(s)}| &\approx& -2sW_\alpha - s^2 N_{\alpha\beta} W^\beta
 +i s^3 G W^2 \bar W_\alpha \nn\\&&
-\frac{s^2}2 \xi_m \gamma^m_{\alpha\beta}\bar W^\beta W^2
+\frac{3s^3}{4}\xi_m N_\alpha^\gamma \gamma^m_{\beta\gamma}\bar
W^\beta W^2
-\frac{5s^3}{12}\xi_m (\gamma^m N)\bar W_\alpha W^2
\nn\\&&
+\frac{is^4}6 G N_{\alpha\beta} \bar W^\beta W^2\,.
\label{PY}
\eea

In a similar way we find
\bea
\bar\nabla_\alpha K_-(z,z'|s)\big|&=&
\frac1{(4\pi i s)^{3/2}}\tilde P_\alpha(s) e^{\tilde Y(s)}  e^{isG^2}e^{\frac
i{4s}\xi^2}\big|\,,\label{DK-}\\
\tilde P_\alpha(s)e^{\tilde Y(s)}\big|&\approx&2s \bar W_\alpha
 +s^2 N_{\alpha\beta}\bar W^\beta
 -is^3 G \bar W^2 W_\alpha
 \nn\\&&
+\frac{s^2}{2}\xi_m \gamma^m_{\alpha\beta} W^\beta \bar W^2
-\frac{3s^3}{4}\xi_m N_\alpha^\gamma \gamma^m_{\gamma\beta}
W^\beta \bar W^2
+\frac{5s^3}{12}\xi_m (\gamma^m N)W_\alpha \bar W^2
\nn\\&&
+\frac{is^4}6 G N_{\alpha\beta} W^\beta \bar W^2\,.
\label{3.44}
\eea

Substituting (\ref{DK+}) and (\ref{DK-}) into (\ref{3.35}) and
using explicit form of the functions (\ref{PY}) and (\ref{3.44})
we perform Gaussian integration over $d^3\xi$,
\bea
\Gamma_{\rm B} &=& \frac{im^2}{16\pi^2 k}
\int d^7z \int_0^\infty\frac{ds\, dt\sqrt{st}}{(s+t)}
e^{i(s+t)(G^2+m^2)}
[-4 W^\alpha \bar W_\alpha
\nn\\&&
+2(s-t)W^\alpha N_{\alpha\beta}\bar W^\beta
+2i(s^2+t^2)GW^2 \bar W^2]\,.
\label{3.45}
\eea
Note that the term containing $W^\alpha \bar W_\alpha$ in
(\ref{3.45}) does not contribute to the effective action according
to (\ref{f-zero}). The first term in the second line of
(\ref{3.45}) also vanishes since it is odd under the change of
integration variables $s$ and $t$. After computing the integrals
over $s$ and $t$ in the last term in (\ref{3.45}) we obtain
\be
\Gamma_{\rm B} = -\frac{15m^2}{256\pi k}\int d^7 z \frac{G\, W^2 \bar
W^2}{(G^2+m^2)^4}\,.
\label{3.46}
\ee

\subsection{Summary of two-loop computations}
\label{sect2.6}
The two-loop low-energy effective action is given by the sum of
eqs.\ (\ref{3.34}) and (\ref{3.46}),
\be
\Gamma^{(2)} = -\frac{15}{256\pi k}\int d^7z \frac{G\,W^2 \bar W^2}{(G^2 +
m^2)^3}\,.
\label{G2fin}
\ee
This expression shows that the functions $f_1(G)$  and $f_2(G)$ in (\ref{eff1})
receive no two-loop quantum corrections,
\be
f_1^{(2)}(G)=f_2^{(2)}(G)=0\,,
\label{2.77}
\ee
and only the function $f_3(G)$ gets non-trivial two-loop
contribution,
\be
f_3^{(2)}(G) =-\frac{15}{256\pi k}\frac{G}{(G^2 +
m^2)^3}\,.
\label{2.78}
\ee

It is instructive to compare the two-loop low-energy effective action
(\ref{G2fin}) with analogous result
in $\cN=2$ SQED with vanishing topological mass considered in \cite{BMS1}.
 The latter is described by the classical action
similar to (\ref{action0}), but in which the gauge
superfield $V$ has $\cN=2$ supersymmetric Maxwell rather than the Chern-Simons term.
The four-derivative low-energy effective action
has the same form (\ref{eff1}), but with the functions $f_i$ given
by (see Appendix \ref{AppC} for details of derivation of these functions)
\bea
\tilde f_1^{(2)} &=&-\frac{g^2}{16\pi^2}\ln(G^2+m^2)\,,\label{tf1}\\
\tilde f_2^{(2)} &=&\frac{5 g^2}{192\pi^2}\frac{G}{(G^2+m^2)^3} \,,\label{tf2}\\
\tilde f_3^{(2)} &=&\frac{g^2}{\pi^2}\frac{98G^2-73 m^2}{3072(G^2+m^2)^4}\,.
\label{tf3}
\eea
Here we put tilde on these functions to distinguish them
from (\ref{2.77}) and (\ref{2.78}).

The obvious difference of the functions $\tilde f_i^{(2)}$ from
$f_i^{(2)}$ is that they contain dimensionful gauge coupling
constant $g^2$. Therefore, even in the massless limit $m=0$, the
functions $\tilde f_i^{(2)}$ give non-conformal effective action
while $f_i^{(2)}$ do.

Let us discuss conformal properties of the effective action
(\ref{G2fin}). Of course, the model (\ref{action0}) is
non-conformal as it explicitly involves the mass parameter $m$,
but we can still get profit from the power of constraints of the
superconformal group either by considering the corresponding massless
theory, $m=0$, or by promoting the mass parameter to a chiral
superfield. The latter option is closer to the $\cN=3$
supersymmetric electrodynamics considered in sect.\ \ref{sect4},
but here, for the sake of simplicity, we will discuss only the massless case,
\be
\Gamma^{(2)}\big|_{m=0} = -\frac{15}{256\pi k}\int d^7z \frac{W^2 \bar W^2}{G^5 }\,.
\label{Gm=0}
\ee
Being scale invariant, this effective action is not $\cN=2$ superconformal
as the superfields $W_\alpha$ and $\bar W_\alpha$ are not
quasi-primaries \cite{BPS1}. The latter means
that these superfields do not have right transformation lows of
superconformal spin-tensors of engineering dimension
3/2.\footnote{The representation of superconformal group in $\cN=2$, $d=3$
superspace was considered in \cite{Park}.}
Nevertheless,
this does not imply any anomaly of the superconformal symmetry.
Recall that the expression (\ref{Gm=0}) was derived for the
background gauge superfield obeying supersymmetric Maxwell
equations (\ref{c1}). Now, one can add some terms with
$D^\alpha W_\alpha$ or $\bar D^\alpha \bar W_\alpha$ to the action
(\ref{Gm=0}) to make it superconformal.\footnote{Similar procedure
was applied in $\cN=2$, $d=4$ superspace to construct superconformal
off-shell effective action for gauge superfield \cite{BKT}.
}

In \cite{BPS1} it was shown that the object
\be
\Psi = \frac iG D^\alpha \bar D_\alpha\ln G
\label{Psi}
\ee
transforms as a scalar with vanishing scaling dimension under
$\cN=2$, $d=3$ superconformal group. Up to a term proportional
to the super Maxwell equations (\ref{c1}), this superfield has the
following expression in terms of the superfield strengths
$W_\alpha$ and $\bar W_\alpha$,
\be
\Psi = -i\frac{W^\alpha \bar W_\alpha}{G^3}\,.
\ee
Hence, the superconformal generalization of the action (\ref{Gm=0})
reads
\be
\Gamma^{(2)}\big|_{m=0}
=\frac{15}{128\pi k}\int d^7z\frac{(D^\alpha \bar D_\alpha \ln G)^2}{G}\,.
\label{Gmconf}
\ee

Representing the action (\ref{Gm=0}) in the superconformal form
(\ref{Gmconf}) has several important consequences. First, we point
out that for the action (\ref{Gmconf}) we can now relax the
constraints (\ref{c1}) on the background gauge superfield which
were used in the derivation of this result. Indeed, the
superconformal invariance allows us to uniquely restore in the
final answer the terms proportional to the supersymmetric Maxwell
equations which were omitted in the intermediate steps of deriving
eq.\ (\ref{Gm=0}).

Second, it is clear now that the function $f_2(G)$
in (\ref{eff1}) should vanish as the corresponding term in the effective action does not
have a superconformal generalization. The unique
superconformal generalization of the four-derivative term is given
by (\ref{Gmconf}) which corresponds to the last term in
(\ref{eff1}).

Finally, it is clear now that it is the superconformal symmetry
which forbids any higher-loop quantum corrections to the function
$f_1(G)$ in (\ref{eff1}). Indeed, the superconformal
generalization of the two-derivative term in the effective action
is given uniquely by $\int d^7z \, G\ln G$, which is nothing but the
one-loop contribution (\ref{f11}) in the massless limit.

Thus, we conclude that the superconformal invariance imposes
strong constraints on the structure of two-loop quantum
corrections to the low-energy effective action (\ref{eff1}) in the model
(\ref{action0}). The similar model with the Maxwell term for the
gauge superfields has no superconformal properties and the
structure of its effective action is much reacher, as is seen in
(\ref{tf1})--(\ref{tf3}).


\section{Generalizations to other Abelian Chern-Simons matter
models}
\subsection{Two-loop effective action in supersymmetric
electrodynamics with one chiral superfield}
\label{sect3}
The results of the previous section can be easily extended to the
Chern-Simons matter model with one chiral superfield,
\be
S=\frac{k}{2\pi}\int d^7 z\, VG - \int d^7 z\, \bar Q
e^{2V} Q\,.
\label{action1}
\ee
This model is known to be superconformal \cite{BPS1}, but has
parity anomaly \cite{NS,Redlich1,Redlich2,AHISS}. The parity
anomaly manifests itself in the presence of the Chern-Simons term
in the one-loop effective action \cite{BPS1},
\be
\Gamma_{\rm odd}=\frac1{4\pi}\int d^7z\, VG\,.
\ee
The subscript ``odd'' here means that the induced Chern-Simons
term is unique part of the effective action which is parity-odd.
This induced Chern-Simons term gives half-integer shift to the
classical value of the Chern-Simons level $k$,
\be
k_{\rm eff}=k+\frac12\,.
\label{k-eff}
\ee
In quantum theory the effective coupling $k_{\rm eff}$ rather than $k$
quantizes in integers, $k_{\rm eff}\in \mathbb
Z$.

The rest of the effective action is parity-even and
we denote it as $\Gamma_{\rm
even}$. So, all the conclusions of sect.\ \ref{general} apply to
it. Hence, its general structure should be the same as of
(\ref{eff1}). The one-loop contributions to the functions $f_i(G)$
for the model (\ref{action1}) were found in \cite{BPS1},
\be
f^{(1)}_1 = \frac1{4\pi} G\ln G\,,\quad
f^{(1)}_2 = 0\,,\quad
f^{(1)}_3 = \frac1{256\pi}\frac1{G^5}\,.
\ee
Our aim now is to compute two-loop corrections to this result,
i.e., to find $f^{(2)}_i$.

The two-loop effective action in the model (\ref{action1}) is
given by the formula
\be
\Gamma^{(2)}= -\int d^7z\, d^7z'
G_{+-}(z,z')G_{-+}(z,z')G(z,z')\,.
\label{GG}
\ee
This effective action corresponds to the first graph in fig.\
\ref{fig1}.

The expression (\ref{GA}) resembles from (\ref{GG}) by the factor $2$. Hence, we
can immediately borrow the result from sect.\
\ref{sectA}: One should divide by two the equation (\ref{3.34})
and apply the massless limit $m\to 0$,
\be
\Gamma^{(2)}= -\frac{15}{512\pi k_{\rm eff}}\int d^7z \frac{ W^2\bar
W^2}{G^5}\,.
\label{3.6}
\ee
Here we also used the effective Chern-Simons level $k_{\rm eff}$
which includes one-loop correction to the classical value,
(\ref{k-eff}).

The effective action (\ref{3.6}) corresponds to the following
values of the functions $f^{(2)}_i$ in (\ref{eff1}),
\be
f^{(2)}_1=f^{(2)}_2=0\,,\qquad
f^{(2)}_3=-\frac{15}{512\pi k_{\rm eff}}\frac1{G^5}\,.
\ee

Since the model (\ref{action1}) is superconformal, the two-loop
effective action (\ref{3.6}) can be represented in a superconformal form.
Similarly as for the action (\ref{Gm=0}), by adding the terms with
$D^\alpha W_\alpha$ and $\bar D^\alpha \bar W_\alpha$, the quantity (\ref{3.6}) can
be recast as follows
\be
\Gamma^{(2)}
=\frac{15}{256\pi k_{\rm eff}}\int d^7z\frac{(D^\alpha \bar D_\alpha \ln G)^2}{G}\,.
\ee
Summarizing now one- and two-loop results, we get the parity-even
part of the two-loop effective action in the superconformal form,
\be
\Gamma_{\rm even}=
\Gamma^{(1)}+\Gamma^{(2)} =\frac1{4\pi}\int d^7 z\, G \ln G
+\frac1{128\pi}\left(\frac{15}{2k_{\rm eff}}-1\right)\int d^7z
\frac{(D^\alpha \bar D_\alpha \ln G)^2}{G}\,.
\label{3.11}
\ee
As is explained in sect.\ \ref{sect2.6}, once the effective action
is represented in the superconformal form (\ref{3.11}), the constraint
(\ref{c1}) can be relaxed. Eq.\ (\ref{3.11}) represents the
parity-even part of the
low-energy effective action in the model (\ref{action1}) up to the
four-derivative order.

The two-loop effective actions obtained in this and previous
sections can be easily generalized to Abelian $\cN=2$ Chern-Simons
matter models with arbitrary number of chiral matter superfields.

\subsection{$\cN=3$ Chern-Simons electrodynamics}
\label{sect4}
The classical action of $\cN=3$ Chern-Simons electrodynamics reads
\bea
\label{action2}
S_{\cN=3} &=& S_{\cN=3}^{\rm CS} + S_{\rm hyper}\,,\\
S_{\cN=3}^{\rm CS}&=& \frac k{2\pi} \int d^7z \, VG
-\frac{ik}{4\pi}\int d^5z \, \Phi^2
-\frac{ik}{4\pi}\int d^5\bar z\, \bar \Phi^2\,,\\
S_{\rm hyper}&=&
- \int d^7 z\,\left( \bar Q_+
e^{2V} Q_+ + \bar Q_- e^{-2V} Q_- \right) - \left( \int d^5z
\,\Phi Q_+Q_-  + {\it c.c.}\right),
\eea
where $\Phi$ is a chiral superfield which is part of the $\cN=3$
gauge multiplet $(V,\Phi)$. Note that this model reduces to
(\ref{action0}) for $\Phi = m$. However, in contrast to
(\ref{action0}), the action of the $\cN=3$
Chern-Simons electrodynamics (\ref{action2}) is superconformal.

Let us make the background-quantum splitting for the $\cN=3$ gauge
multiplet,
\be
(V,\Phi) \to (V,\Phi) + (v,\phi)\,,
\ee
where the superfields $(V,\Phi)$ in the r.h.s.\ are treated as
background while $(v,\phi)$ as the quantum ones. Under this
splitting the part of the $\cN=3$ Chern-Simons action which is
quadratic with respect to the quantum superfields reads
\be
S_{\cN=3}^{\rm CS}= \frac{ik}{4\pi}\left(\int d^7z \,
vD^\alpha \bar D_\alpha v
-\int d^5z\, \phi^2 -\int d^5\bar z \, \bar\phi^2
\right)+\ldots,
\label{Sdots}
\ee
where dots stand for the linear terms for the quantum superfields
which are irrelevant in quantum loop computations.
Note that the superfield $\phi$ is gauge invariant since the gauge
group is Abelian. Hence, to fix the gauge freedom it is sufficient to add
to (\ref{Sdots}) the same gauge fixing term (\ref{Sgf}) as in the
$\cN=2$ case. This yields the following action for quantum
superfields up to quartic order,
\bea
S_{\rm quant}&=&S_2+S_{\rm int}\,,\\
S_2&=&\int d^7z\left(\frac{ik}{4\pi}vHv
-\bar {\cal Q}_+ {\cal Q}_+ -\bar{\cal Q}_-{\cal Q}_- \right)
\nn\\&&
-\int d^5z\left(\frac{ik}{4\pi}\phi^2 +\Phi{\cal Q}_+ {\cal Q}_-\right)
-\int d^5\bar z \left(\frac{ik}{4\pi}\bar\phi^2
-\bar\Phi\bar{\cal Q}_+ \bar{\cal Q}_-
\right),\\
S_{\rm int}&=&-2\int d^7z\left[
(\bar{\cal Q}_+ {\cal Q}_+ - \bar{\cal Q}_- {\cal Q}_-)v
+(\bar{\cal Q}_+{\cal Q}_+  +\bar{\cal Q}_- {\cal Q}_-)v^2
\right]\nn\\&&
-\int d^5z \, \phi{\cal Q}_+ {\cal Q}_-
+\int d^5\bar z\, \bar\phi \bar{\cal Q}_+ \bar{\cal Q}_-
+O(v^3)\,.
\eea

\begin{figure}[t]
\begin{center}
\setlength{\unitlength}{0.9mm}
\begin{picture}(100,50)
\thicklines
\qbezier(35,25)(35,30.74)(30.6,35.6)
\qbezier(30.6,35.6)(25.75,40)(20,40)
\qbezier(20,40)(14.26,40)(9.39,35.6)
\qbezier(9.39,35.6)(5,30.74)(5,25)
\qbezier(5,25)(5,19.26)(9.39,14.39)
\qbezier(9.39,14.39)(14.26,10)(20,10)
\qbezier(20,10)(25.74,10)(30.6,14.39)
\qbezier(30.6,14.39)(35,19.26)(35,25)
\put(5,25){\circle*{2}}
\put(35,25){\circle*{2}}
\put(2,35){${\cal Q}_+$}
\put(32,35){${\cal Q}_-$}
\put(2,13){${\cal Q}_-$}
\put(32,13){${\cal Q}_+$}
\put(8,27){$\phi$}
\put(30,27){$\phi$}
\qbezier(5,25)(6,26)(6,26)
\qbezier(6,26)(7,25)(8,24)
\qbezier(8,24)(9,25)(10,26)
\qbezier(10,26)(11,25)(12,24)
\qbezier(12,24)(13,25)(14,26)
\qbezier(14,26)(15,25)(16,24)
\qbezier(16,24)(17,25)(18,26)
\qbezier(18,26)(19,25)(20,24)
\qbezier(20,24)(21,25)(22,26)
\qbezier(22,26)(23,25)(24,24)
\qbezier(24,24)(25,25)(26,26)
\qbezier(26,26)(27,25)(28,24)
\qbezier(28,24)(29,25)(30,26)
\qbezier(30,26)(31,25)(32,24)
\qbezier(32,24)(33,25)(34,26)
\qbezier(34,26)(35,25)(35,25)
%
%
%
%
\qbezier(95,25)(95,30.74)(90.6,35.6)
\qbezier(90.6,35.6)(85.75,40)(80,40)
\qbezier(80,40)(74.26,40)(69.39,35.6)
\qbezier(69.39,35.6)(65,30.74)(65,25)
\qbezier(65,25)(65,19.26)(69.39,14.39)
\qbezier(69.39,14.39)(74.26,10)(80,10)
\qbezier(80,10)(85.74,10)(90.6,14.39)
\qbezier(90.6,14.39)(95,19.26)(95,25)
\put(65,25){\circle*{2}}
\put(95,25){\circle*{2}}
\put(62,35){$\bar{\cal Q}_+$}
\put(92,35){$\bar{\cal Q}_-$}
\put(62,13){$\bar{\cal Q}_-$}
\put(92,13){$\bar{\cal Q}_+$}
\put(68,27){$\bar\phi$}
\put(90,27){$\bar\phi$}
\qbezier(65,25)(66,26)(66,26)
\qbezier(66,26)(67,25)(68,24)
\qbezier(68,24)(69,25)(70,26)
\qbezier(70,26)(71,25)(72,24)
\qbezier(72,24)(73,25)(74,26)
\qbezier(74,26)(75,25)(76,24)
\qbezier(76,24)(77,25)(78,26)
\qbezier(78,26)(79,25)(80,24)
\qbezier(80,24)(81,25)(82,26)
\qbezier(82,26)(83,25)(84,24)
\qbezier(84,24)(85,25)(86,26)
\qbezier(86,26)(87,25)(88,24)
\qbezier(88,24)(89,25)(90,26)
\qbezier(90,26)(91,25)(92,24)
\qbezier(92,24)(93,25)(94,26)
\qbezier(94,26)(95,25)(95,25)
\put(43,3){Type C}
\put(49,24){$+$}
  \end{picture}
\end{center}
\caption[b]{Two-loop supergraphs in $\cN=3$ supersymmetric electrodynamics
which involve (anti)chiral propagators $\langle \phi \phi \rangle$ and
$\langle \bar\phi \bar \phi \rangle$.}
\label{fig2}
\end{figure}
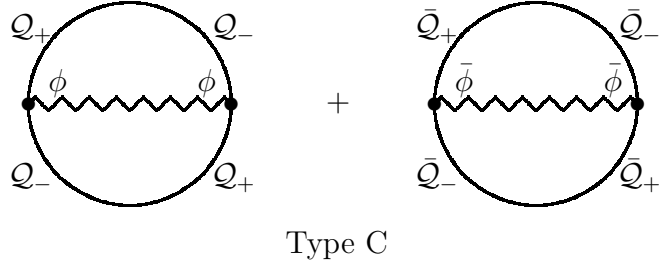

The action $S_{\rm int}$ is responsible for the interaction vertices
while $S_2$ gives propagators for the quantum superfields. As
compared with the $\cN=2$ electrodynamics, there is a new vertex
$\phi{\cal Q}_+ {\cal Q}_-$ (and its conjugate) and the propagators
$\langle\phi \phi \rangle$, $\langle \bar\phi \bar\phi \rangle$,
\be
\langle\phi(z)\phi(z')\rangle =
-\frac{2\pi}{k}\delta_+(z,z')\,,\qquad
\langle\bar\phi(z)\bar\phi(z')\rangle =
-\frac{2\pi}{k}\delta_-(z,z')\,.
\label{phi-prop}
\ee
Hence, apart from the graphs in fig.\ \ref{fig1},
there are two extra Feynman graphs in the $\cN=3$ SQED with these
propagators which are represented in fig.\ \ref{fig2}.
Correspondingly, two-loop effective action is given by the following formal
expression
\bea
\Gamma^{(2)}_{\cN=3}&=&\Gamma_{\rm A}+\Gamma_{\rm B}+\Gamma_{\rm C}\,,\\
\Gamma_{\rm A}&=& -2\int d^7z\, d^7z'
G_{+-}(z,z')G_{-+}(z,z')G(z,z')\,,
\\
\Gamma_{\rm B}&=& -2\int d^7z\, d^7z'\,\Phi\bar\Phi
 G_+ (z,z')G_-(z,z')G(z,z')\,,
\\
\Gamma_{\rm C}&=&\frac{\pi i}k\int d^5z \,
G_+(z,z')G_+(z',z)\delta_+(z,z') + c.c.
\label{GC}
\eea
The chiral delta-function in the expression (\ref{GC})
originates from the propagators (\ref{phi-prop}).

Recall that the background gauge superfield $V$ is constrained by
(\ref{c1}) and (\ref{c2}). Analogous constraints for $\Phi$,
\be
D_\alpha\Phi=0\,,\qquad
\bar D_\alpha\bar\Phi=0\,,
\label{chir-back}
\ee
just mean that this superfield is simply a constant. For such a background the
heat kernels for the propagators $G_{+-}$ and $G_+$ are given
in Appendix \ref{AppB}. In particular, the equation
(\ref{K+limit}) shows that at coincident superspace points the
heat kernel $K_+$ is proportional to $W^2$,
\be
K_+(z,z|s)\propto W^2\,.
\ee
The quantity (\ref{GC}) contains two chiral propagators $G_+$ at
coincident superspace points after integration over $dz'$ with
the help of chiral delta-function. Hence, $\Gamma_{\rm C}$
vanishes as it contains too many $W$'s,
\be
\Gamma_{\rm C}=0\,.
\ee

It is clear that for the constant chiral superfield background
(\ref{chir-back}) computations of the contributions $\Gamma_{\rm
A}$ and $\Gamma_{\rm B}$ to the two-loop effective action are
absolutely identical to the ones given in sections \ref{sectA}
and \ref{SectB}. Hence, we can borrow the result (\ref{G2fin})
just by promoting the mass parameter to the chiral superfield,
\be
\Gamma^{(2)}_{\cN=3} = -\frac{15}{256\pi k}\int d^7z \frac{G\,W^2 \bar W^2}{(G^2 +
\Phi\bar\Phi)^3}\,.
\label{G2finN3}
\ee

The effective action (\ref{G2finN3}) is scale invariant, but is
not superconformal similarly as the effective action (\ref{3.6})
obtained in the previous section. To construct a superconformal
generalization of (\ref{G2finN3}) we use a version of the
quasi-primary superfield (\ref{Psi}) which involves the chiral
superfield $\Phi$ \cite{BPS1},
\be
{\bf \Psi}=\frac iG D^\alpha\bar D_\alpha \ln
(G+\sqrt{G^2+\Phi\bar\Phi})\,.
\ee
Up to a term proportional to the super Maxwell equations
(\ref{c1}), this superfield reads
\be
{\bf \Psi}=-i\frac{W^\alpha\bar
W_\alpha}{(G^2+\Phi\bar\Phi)^{3/2}}\,.
\ee
Hence, the superconformal generalization of (\ref{G2finN3}) is
given by
\be
\Gamma^{(2)}_{\cN=3} =-\frac{15}{128 \pi k}
\int d^7z\, G{\bf \Psi}^2
=\frac{15}{128\pi k}\int d^7z\frac1G
[D^\alpha\bar D_\alpha \ln
(G+\sqrt{G^2+\Phi\bar\Phi})]^2\,.
\label{N3eff2}
\ee
The representation of the effective actions (\ref{N3eff2}) in
superconformal form allows us to relax the constraint (\ref{c1})
which was used for deriving this result.

For completeness, we present here the four-derivative part of
one-loop effective action in the model (\ref{action2})
which was found in \cite{BPS1}:
\be
\Gamma^{(1)}_{\cN=3}=\frac1{64\pi}\int d^7z \, {\bf\Psi}^2\sqrt{G^2+\Phi\bar\Phi}
=-\frac1{64\pi}\int d^7z\frac{\sqrt{G^2+\Phi\bar\Phi}}{G^2}
[D^\alpha\bar D_\alpha \ln
(G+\sqrt{G^2+\Phi\bar\Phi})]^2\,.
\label{N3eff1}
\ee
It is interesting to note that the expressions (\ref{N3eff2}) and
(\ref{N3eff1}) have slightly different functional structure. This
is explained by the fact that the two-loop effective action
(\ref{G2finN3}) was obtained in the gauge (\ref{gauge}) which is only $\cN=2$
supersymmetric. As a consequence, the two-loop result
(\ref{G2finN3}) does not respect full $\cN=3$ superconformal group
and requires $\cN=3$ supersymmetrization. The issue of finding
$\cN=3$ supersymmeteric versions of the actions (\ref{N3eff2})
and (\ref{N3eff2}) deserves a separate study.

The most natural way to obtain the effective action in the model
(\ref{action2}) in explicitly $\cN=3$ supersymmetric form is by
using the $\cN=3$, $d=3$ harmonic superspace
\cite{Zupnik1,Zupnik2}. Quantum aspects of supersymmetric gauge
theories in this superspace were studied in \cite{BILPSZ}. It
would be interesting to explore the low-energy effective action in
$\cN=3$ gauge theories using this approach.

\section{Conclusions}

Recently, we computed two-loop low-energy effective actions in the
$\cN=2$ and $\cN=4$ SQED theories \cite{BMS1} with vanishing topological mass.
In the present paper we
considered similar models in which the gauge superfield is
described by the Chern-Simons rather than the supersymmetric
Maxwell action. In these models we computed two-loop
low-energy effective actions up to four-derivative order in the gauge superfield sector and
compared them with similar results in the SQED theories considered
in \cite{BMS1}. In the massless case these
Chern-Simons matter models are superconformal. We demonstrated that
the superconformal invariance imposes strong restrictions on the
structure of two-loop effective actions forbidding a number of
superfield structures (described by the functions $f_1$ and $f_2$ in (\ref{eff1}))
 which are non-trivial in similar SQED
theories with vanishing topological mass. Note that any superconformal effective action for the $\cN=2$
gauge superfield can be expressed in terms of superconformal
invariants classified in \cite{BPS1}. So, the quantum loop
computations performed in the present paper
only fix numerical coefficients in the decomposition
of the effective action over these invariants.

The low-energy effective action in the $\cN=3$ Chern-Simons
electrodynamics is also expressed in terms of $\cN=2$
superconformal invariants. However, the full $\cN=3$ supersymmetry
is not explicit as the two-loop effective action is computed in
the $\cN=2$ supersymmetric gauge. The most natural way of
recasting this effective action in the $\cN=3$ supersymmetric form
is based on the $\cN=3$, $d=3$ harmonic superspace
\cite{Zupnik1,Zupnik2}. Some quantum aspects of supersymmetric gauge
theories in this superspace were studied in \cite{BILPSZ}. It
would be interesting to explore the low-energy effective action in
$\cN=3$ gauge theories using this approach.

The results of the present paper, together with similar results of
\cite{BMS1}, give the structure of low-energy effective actions in
Abelian three-dimensional $\cN=2$ and $\cN=3$ supersymmetric gauge
theories in two particular cases, when the gauge superfield is
described either by Chern-Simons or by pure super Maxwell action. The latter
corresponds to vanishing topological mass while the former
describes gauge superfield with infinitely large topological mass.
It would be
interesting to consider more general case of the supersymmetric
gauge theories with a finite value of the topological mass. The
effective actions in such models should interpolate between the
results of the present paper and those of \cite{BMS1}. Another
natural generalization could be a computation of two-loop quantum
corrections to low-energy effective actions in non-Abelian gauge
theories in the $\cN=2$, $d=3$ superspace considered, e.g., in
\cite{BPS2}.

In the present paper we studied the effective action in the gauge
superfield sector. It is interesting to consider also the part of
the effective action for (anti)chiral superfields and, in
particular, to study two-loop effective K\" ahler potential.
In components, such an effective action is responsible, in particular, for the
effective scalar potential. This problem was
studied for $\cN=1$, $d=3$ superfield models in
\cite{Lehum1,Lehum2,Lehum3,Lehum4} and for pure $\cN=2$, $d=3$
Wess-Zumino model in \cite{BMS}. It is natural to extend the results of the
latter work to models of $\cN=2$ and $\cN=3$ SQED considered in the present
paper and compare them with analogous results for the $\cN=1$ models.
In non-supersymmetric three-dimensional scalar electrodynamics the two-loop
effective potential was studied in \cite{Tekin1,Tekin2}.

Finally, it is very tempting to study the structure of low-energy
effective actions in the BLG and ABJM models. This problem becomes
very hot in the light of recent discussion in \cite{Schwarz} where
the relations of such an effective action to the dynamics of M2
branes was proposed. We expect that the techniques of quantum
computations in the $\cN=2$, $d=3$ superspace developed in
\cite{BPS1,BPS2,BPS-back,BMS1} and in the present paper might be
useful for studying this issue. Alternatively, the $\cN=3$
harmonic superspace formulation \cite{N3ABJM} of the ABJM and BLG
models can be employed.

\vspace{5mm}
{\bf Acknowledgments.} I.B.S. wishes to thank D.~Sorokin and
N.~Pletnev for stimulating discussions and comments. The work was
partially supported by the RFBR grant Nr.\ 12-02-00121 and by LRSS
grant Nr.\ 88.2014.2. I.L.B.\ and I.B.S.\ acknowledge the support
from RFBR grants Nr.\ 13-02-90430 and 13-02-91330 and DFG grant LE
838/12/1. I.L.B.\ and B.S.M.\ are thankful to the grant of Russian
Ministry of Education and Science, project \ TSPU-122 for partial
support. The work of I.B.S.\ was also supported by the Marie Curie
research fellowship Nr.\ 909231 ``Quantum Supersymmetry'' and by the
Padova University Project CPDA119349. The work of B.S.M\ was partly
supported by the RFBR grant for young researchers Nr. 14-02-31201.

\appendix
\section{Parallel displacement propagator in $\cN=2$, $d=3$
superspace}
\label{appA}
The technique of gauge-covariant multiloop quantum computations in
$\cN=1$, $d=4$ superspace was developed in \cite{Kuz03}. Its power
was demonstrated in studying two-loop effective actions in the
$\cN=1$ and $\cN=2$, four-dimensional SQEDs in
\cite{Kuz0310,Kuz07} and other gauge theories with extended
supersymmetry in $\cN=1$ superspace,
\cite{Kuzenko:2007tf,Kuzenko:2004ma,Kuzenko:2004yd,Kuzenko:2003wu}.

 The key
ingredient of this technique is the parallel displacement
propagator $I(z,z')$ which relates gauge-covariant objects in
different superspace points. In the $\cN=2$, $d=3$ superspace the
parallel displacement propagator was considered in \cite{BMS1}.
Here we review basic properties of this object which are necessary
for two-loop quantum computations in the $\cN=2$
Chern-Simons matter model studied in this paper.

The parallel displacement propagator $I(z,z')$ is a two-point
superspace function taking its values in the gauge group and
depending on the gauge superfields
with the following properties:
\begin{itemize}
\item[(i)]
Under gauge transformations it changes as
\be
I(z,z')\to e^{i\tau(z)}I(z,z')e^{-i\tau(z')}\,,
\label{prop1}
\ee
with $\tau(z)$ being a real gauge superfield parameter;
\item[(ii)]
It obeys the equation
\be
\zeta^A \nabla_A I(z,z')=\zeta^A\left(D_A +
V_A(z)\right)I(z,z')=0\,,
\label{prop2}
\ee
where $V^A$ are gauge connections for $D^A$ and
$\zeta^A=(\xi^m,\zeta^\alpha,\bar\zeta_\alpha)$ is the
$\cN=2$ supersymmetric interval,
\be
\zeta^\alpha=(\theta-\theta')^\alpha\,,\quad
\bar\zeta^\alpha= (\bar\theta-\bar\theta')^\alpha\,,\quad
\xi^m=(x-x')^m-i\gamma^m_{\alpha\beta}\zeta^\alpha \bar\theta'^\beta
+i\gamma^m_{\alpha\beta}\theta'^\alpha \bar\zeta^\beta\,;
\label{superinterval}
\ee
\item[(iii)] For coincident superspace points $z=z'$ it reduces to the
identity operator in the gauge group,
\be
I(z,z)=1\,.
\label{prop3}
\ee
\end{itemize}

The properties (\ref{prop1})--(\ref{prop3}) allow one to express the
derivatives of the parallel transport propagator in terms of the parallel transport
propagator itself and gauge-covariant superfield strengths. In
particular, the following equations hold \cite{BMS1}:
\bea
\nabla_\beta I(z,z')&=&\bigg[
-i\bar\zeta_\beta G +\frac12\xi_{\alpha\beta} \bar W^\alpha-\frac
i{12}\bar\zeta^2 W_\beta +\frac i6\bar\zeta_\beta\zeta^\alpha \bar
W_\alpha -\frac i3 \bar\zeta^\alpha \zeta_\alpha \bar W_\beta
\nn\\
&&+\frac1{12} \bar \zeta^\alpha \xi_{\beta\gamma}
\bar \nabla_\alpha \bar W^\gamma
-\frac1{12} \bar\zeta^\alpha \xi_{\alpha\gamma}
\bar\nabla^\gamma \bar W_\beta
-\frac i{12}\bar\zeta^2 \zeta_\beta \bar\nabla^\alpha \bar W_\alpha
\bigg]I(z,z')\,,\label{DI1}\\
\bar\nabla^\beta I(z,z')&=&\bigg[
-i \zeta^\beta G - \frac 12 \xi_{\alpha}^\beta W^\alpha
+\frac i{12} \zeta^2 \bar W^\beta
- \frac i6 \zeta^\beta \bar\zeta^\alpha W_\alpha
+\frac i3 \zeta^\alpha \bar \zeta_\alpha W^\beta
\nn\\&&
+\frac1{12}\zeta_\alpha \xi^{\beta\gamma}\nabla^\alpha W_\gamma
-\frac1{12}\zeta_\alpha  \xi^{\alpha\gamma} \nabla_\gamma W^\beta
-\frac i{12}\zeta^2 \bar\zeta^\beta \nabla^\alpha
W_\alpha\bigg]I(z,z')\,.
\label{DI2}
\eea

\section{Green's functions in $\cN=2$, $d=3$ superspace}
\label{AppB}
Consider a covariantly chiral superfield $\Phi$, $\bar\nabla_\alpha
\Phi=0$, where $\nabla_\alpha$ and $\bar\nabla_\alpha$ are
gauge-covariant spinor derivatives. There are two types of Green's
functions for this superfield: $G_+(z,z')$ which is chiral with
respect to both arguments and $G_{+-}(z,z')$ which is chiral
with respect to $z$ and is antichiral with respect to $z'$,
\be
i\langle\Phi(z)\Phi(z')\rangle \equiv m G_+(z,z')\,,\qquad
i\langle\Phi(z)\bar\Phi(z')\rangle \equiv G_{+-}(z,z')\,.
\ee
By definition, they obey the following equations:
\bea
(\square_+ + m^2 )G_+(z,z') &=& -\delta_+(z,z')\,,\\
(\square_- + m^2 )G_-(z,z') &=& -\delta_-(z,z')\,,\\
\frac14\bar \nabla^2 G_{-+}(z,z') + m^2 G_+(z,z') &=&
-\delta_+(z,z')\,,\\
\frac14\nabla^2 G_{+-}(z,z') + m^2 G_-(z,z')&=&
-\delta_-(z,z')\,,\label{A5}
\eea
where $\delta_\pm(z,z')$ are (anti)chiral delta-functions and the
operators $\square_\pm$ are given by
\bea
\square_+&=&\nabla^m\nabla_m+G^2+\frac{i}{2}(\nabla^\alpha W_\alpha) +i
W^\alpha\nabla_\alpha\,,\nn\\
\square_-&=&\nabla^m \nabla_m +G^2 - \frac i2 (\bar\nabla^\alpha \bar W_\alpha)
-i\bar W^\alpha \bar \nabla_\alpha\,.
\eea

It is convenient to express the Green's functions in terms of
corresponding heat kernels,
\bea
G_{\pm}(z,z') &=& i\int_0^\infty ds \, K_\pm(z,z'|s)e^{is\, m^2}\,,
\label{B7}\\
G_{+-}(z,z') &=& i\int_0^\infty ds\, K_{+-}(z,z'|s)e^{is\,
m^2}\,,\label{B8}\\
G_{-+}(z,z') &=& i\int_0^\infty ds\, K_{-+}(z,z'|s)e^{is\, m^2}\,.
\label{B9}
\eea
Explicit expressions for these heat kernels were found in
\cite{BMS1}:
\bea
K_+(z,z'|s)&=&
\frac{1}{8(i\pi s)^{3/2}}\frac{sB}{\sinh(sB)}e^{isG^2}
\nn\\&&\times
{\cal O}(s)e^{\frac{i}{4}(F\coth(sF))_{mn}\xi^m\xi^n-
\frac12\bar\zeta^\beta \xi_{\beta\gamma}W^\gamma} \zeta^2
I(z,z')\,,
\label{K_+}
\\
K_-(z,z'|s)&=&
\frac{1}{8(i\pi s)^{3/2}}\frac{sB}
{\sinh(sB)}e^{isG^2}\nn\\&&\times
{\cal O}(s)e^{\frac{i}{4}(F\coth(sF))_{mn}\xi^m\xi^n-
\frac12\zeta^\beta \xi_{\beta\gamma}\bar W^\gamma} \bar\zeta^2
I(z,z')\,,
\label{K_-}
\\
K_{+-}(z,z'|s)&=&-\frac1{8(i\pi s)^{3/2}}\frac{sB}{\sinh(sB)}
e^{isG^2}{\cal O}(s)
e^{\frac i4(F\coth(sF))_{mn}\rho^m\rho^n +R(z,z')}
I(z,z')\,,
\label{K_-+1}\\
K_{-+}(z,z'|s)&=&-\frac1{8(i\pi s)^{3/2}}\frac{sB}{\sinh(sB)}
e^{isG^2}{\cal O}(s)
e^{\frac i4(F\coth(sF))_{mn}\tilde\rho^m\tilde\rho^n +\tilde R(z,z')}
I(z,z')\,.
\label{K_-+2}
\eea
Here $B^2= \frac12N^\alpha_\beta N^\beta_\alpha$ and
${\cal O}(s)$ is the operator of the form
\be
{\cal
O}(s)=
e^{s(\bar W^\alpha \bar\nabla_\alpha - W^\alpha \nabla_\alpha)}\,.
\ee
The functions $R(z,z')$ and $\tilde R(z,z')$ read
 \bea
R(z,z')&=&-i\zeta \bar \zeta G
+\frac{7 i}{12}\bar \zeta^2 \zeta W
+\frac{i}{12} \zeta^2 \bar\zeta \bar W
-\frac12  \bar\zeta^\alpha \rho_{\alpha\beta} W^\beta
-\frac12  \zeta^\alpha \rho_{\alpha\beta} \bar W^\beta
\nn\\&&
+\frac1{12} \zeta^\alpha\bar\zeta^\beta
 [\rho_\beta^\gamma D_\alpha  W_\gamma
 -7\rho_\alpha^\gamma  D_\gamma  W_\beta]\,,
 \label{R}\\
\tilde R(z,z')&=&i\zeta \bar \zeta G
+\frac i{12}\bar \zeta^2 \zeta W
+\frac{7i}{12} \zeta^2 \bar\zeta \bar W
-\frac12  \bar\zeta^\alpha \tilde \rho_{\alpha\beta}  W^\beta
-\frac12  \zeta^\alpha \tilde \rho_{\alpha\beta} \bar W^\beta
\nn\\&&
+\frac1{12} \zeta^\alpha\bar\zeta^\beta
 [7\tilde \rho_{\beta\gamma}\bar D^\gamma \bar W_\alpha
 -\tilde \rho_{\alpha\gamma}\bar D_\beta \bar W^\gamma]\,.
\label{tildeR}
 \eea
The objects $\rho^m$ and $\tilde\rho^m$ are versions of bosonic
interval $\xi^m$ with specific chirality properties:
\bea&&
\rho^{m}=
\xi^m+i \zeta^\alpha \gamma^m_{\alpha\beta}\bar\zeta^\beta\,,
\qquad
D'_\alpha \rho^m = \bar D_\alpha \rho^m =0\,,
\nn\\&&
\tilde\rho^{m}=
\xi^m-i \zeta^\alpha \gamma^m_{\alpha\beta}\bar\zeta^\beta\,,
\qquad
\bar D'_\alpha \tilde\rho^m =  D_\alpha \tilde\rho^m =0\,.
\label{chirho}
\eea

To make the heat kernels (\ref{K_+}) and (\ref{K_-+1}) more useful
for loop quantum computations one has to push the operator ${\cal
O}(s)$ on the right and act with it on the parallel transport
propagator. The result of this procedure is \cite{BMS1}
\bea
K_+(z,z'|s)&=&
\frac{1}{8(i\pi s)^{3/2}}\frac{sB}
{\sinh(sB)}e^{isG^2}e^{\frac{i}{4}(F\coth(sF))_{mn}\xi^m(s)\xi^n(s)-
\frac12\bar\zeta^\beta(s) \xi_{\beta\gamma}(s)W^\gamma(s)}
\nn\\
&&\times e^{\int_0^s dt \Sigma(z,z'|t)}\zeta^2(s)
I(z,z')\,,
\label{K+fin}\\
K_-(z,z'|s)&=&
\frac{1}{8(i\pi s)^{3/2}}\frac{sB}
{\sinh(sB)}e^{isG^2}e^{\frac{i}{4}(F\coth(sF))_{mn}\xi^m(s)\xi^n(s)
-\frac12\zeta^\beta(s) \xi_{\beta\gamma}(s)\bar W^\gamma(s)}
\nn\\
&&\times e^{\int_0^s dt \Sigma(z,z'|t)}\bar\zeta^2(s) I(z,z')\,,
\label{K-fin}\\
K_{+-}(z,z'|s)&=&-\frac1{8(i\pi s)^{3/2}}\frac{sB}{\sinh(sB)}
e^{isG^2}\nn\\&&\times
e^{\frac i4(F\coth(sF))_{mn}\rho^m(s)\rho^n(s) +R(z,z')
+\int_0^s dt(R'(t)+\Sigma(t))}
I(z,z')\,,
\label{K_-+2_}\\
K_{-+}(z,z'|s)&=&-\frac1{8(i\pi s)^{3/2}}\frac{sB}{\sinh(sB)}
e^{isG^2}
\nn\\&&\times
e^{\frac i4(F\coth(sF))_{mn}\tilde\rho^m(s)\tilde\rho^n(s) +\tilde R(z,z')
+\int_0^s dt(\tilde R'(t)+\Sigma(t))}
I(z,z')\,.
\label{K_-+3_}
\eea
All $s$-dependent objects in these expressions are defined by the
rule $X(s)= {\cal O}(s)X{\cal O}(-s)$, e.g.
\bea
W^\alpha(s)&\equiv&{\cal O}(s)W^\alpha {\cal O}(-s)
=W^\beta (e^{-sN})_\beta{}^\alpha\,,\nn\\
\zeta^\alpha(s)&\equiv&{\cal O}(s)\zeta^\alpha {\cal O}(-s)
= \zeta^\alpha+W^\beta
((e^{-sN}-1)N^{-1})_\beta{}^\alpha\,,\nn\\
\bar\zeta^\alpha(s)&\equiv&{\cal O}(s)\bar\zeta^\alpha {\cal O}(-s)
= \bar\zeta^\alpha+\bar W^\beta
((e^{-s N}-1)N^{-1})_\beta{}^\alpha\,,\nn\\
\xi^m(s)&\equiv&{\cal O}(s)\xi^m{\cal O}(-s)
=\xi^m-i(\gamma^m)^{\alpha\beta}\int_0^s dt\left(W_{\alpha}(t)\bar\zeta_{\beta}(t)
+\bar W_{\alpha}(t)\zeta_{\beta}(t)\right)\,.
\label{id's}
\eea
The quantities $\Sigma(z,z')$ and $R'(z,z')+\Sigma(z,z')$ in (\ref{K+fin})--(\ref{K_-+3_})
are given by
 \bea
\Sigma(z,z')&=&-i(\bar W^\beta \zeta_\beta - W^\beta \bar \zeta_\beta)G
-\frac i3 \zeta^\alpha \bar \zeta^\beta W_\beta \bar W_\alpha
+\frac {2i}3 \zeta^\alpha \bar\zeta_\alpha W^\beta \bar W_\beta
\nn\\&&
+\frac i{12}\zeta^2 [\bar W^2-\bar\zeta^\alpha\bar W_\alpha D^\beta W_\beta]
+\frac i{12}\bar\zeta^2[W^2+\zeta^\alpha W_\alpha \bar D^\beta \bar W_\beta]
\nn\\&&
+\frac1{12}(\zeta^\alpha \bar W^\beta -\bar\zeta^\beta W^\alpha)
[\xi_{\alpha\gamma} D^\gamma W_\beta+\xi_{\beta\gamma}
 \bar D^\gamma \bar W_{\alpha}]\,,
 \label{Sigma}\\
R'+\Sigma&=&
2i\bar\zeta W G +2i( \zeta\bar \zeta\, W\bar W- \zeta W \,
\bar\zeta \bar W)\nn
\\&&
+i\bar\zeta^2[W^2 - \zeta^\alpha W^\beta  D_\alpha W_\beta]
-\frac12\bar\zeta^\beta  W^\alpha
[\rho_{\beta\gamma}\bar D^\gamma \bar W_\beta
- \rho_{\alpha\gamma}  D^\gamma W_\beta]\,,
\label{RS}\\
\tilde R'+\Sigma&=&
-2i\zeta\bar W G +2i( \zeta\bar \zeta\, W\bar W- \zeta W \,
\bar\zeta \bar W)\nn\\&&
+i\zeta^2[\bar W^2 + \bar\zeta^\alpha \bar W^\beta \bar D_\alpha \bar W_\beta]
-\frac12 \zeta^\beta \bar W^\alpha
 [\tilde\rho_{\alpha\gamma}\bar D^\gamma \bar W_\beta
  +\tilde\rho_{\beta\gamma} \bar D_\alpha \bar W^\gamma]\,.
 \eea

The heat kernels (\ref{K_+}) and (\ref{K_-+1}) at coincident
Grassmann superspace points reduce to the following expressions
\cite{BMS1}:
\bea
K_{+}(z,z'|s)\Big|&=&\frac{1}{4(i\pi s)^{3/2}}\frac{s
W^2}B \tanh \frac{sB}{2} e^{is G^2} e^{\frac i4(F \coth (sF))_{mn}\xi^m \xi^n} \,.
\label{K+limit}\\
K_{+-}(z,z'|s)\Big|&=&-\frac{1}{8(i\pi s)^{3/2}}\frac{sB}
{\sinh(sB)}\,e^{isG^2}
\exp\Big\{\frac{i}{4}(F\coth(sF))_{mn}\,\rho^m\rho^n \nn\\
&&-iG W^\alpha f_{\alpha}{}^{\beta}(s) \bar W_\beta+W^\alpha \rho_m f^m_{\alpha\beta}(s)
 \bar W^\beta
+\frac{i}{2} W^2\bar W^2 f(s)
\bigg\}\,,\label{kerlim}
\eea
where
 \bea
f_{\alpha}{}^{\beta}(s)&=&2 B^{-2}(1-sN - e^{-sN})_{\alpha}{}^{\beta}\,, \nn\\
f(s)&=&\frac{1}{sB^4} \bigg[(sB)^2
-4\sinh^2(sB/2)\big(1+sB\tanh(sB/2)\big)\bigg]\,,\nn\\
f^m_{\alpha\beta}(s)&=&\frac12 B^{-2}(\cosh(sB)-1) \bigg[
(e^{-sN})_\beta{}^\gamma N_\alpha{}^\delta\,(\gamma^m)_{\gamma\delta}+
(N(e^{-sN}))_\beta{}^\delta\,(\gamma^m)_{\alpha\delta} \bigg]-    \label{fun}\\
& -& \frac{1}{2}(F\coth(sF))^m{}_n\gamma^n_{\gamma\delta}\bigg[
\big(\frac{e^{-s N}-1}{N}\big)_\alpha{}^\gamma\, \big(\frac{e^{- s
N}-1}{N}\big)_\beta{}^\delta
+\frac{\varepsilon_{\alpha\beta}N^{\gamma\delta}}{B^{3}}(sB-\sinh(sB)) \bigg]\,. \nn
\eea

\section{Two-loop effective action in $\cN=2$ SQED up to four-derivative order}
\label{AppC}
Classical action of $\cN=2$ SQED has the form similar to (\ref{action0}), but
the gauge superfield is described by supersymmetric Maxwell rather
than the Chern-Simons term. The two-loop Euler-Heisenberg effective
action in this model was studied in \cite{BMS1}. In components,
such an action contains all powers of the Maxwell field strength.
Here we wish to consider only the superfield terms up to four-derivative
order, $F^4$, to compare them with the similar ones in the model
(\ref{action0}) studied in section \ref{sect2}. In principle,
these terms can be extracted from the results
obtained in \cite{BMS1} which include all powers of $F_{mn}$ in
components. However, we give here some details of
deriving these terms ``from scratch'', following the same procedure
as in section \ref{sect2} for similar Chern-Simons matter
model (\ref{action0}).

Two-loop effective action in the $\cN=2$ SQED has the structure
analogous to (\ref{Gam}), but with $\Gamma_{\rm A}$ and $\Gamma_{\rm
B}$ given by
\bea
\Gamma_{\rm A}&=& -2g^2\int d^7z\, d^7z'
G_{+-}(z,z')G_{-+}(z,z')G_0(z,z')\,,
\label{GA1}\\
\Gamma_{\rm B}&=& -2g^2m^2\int d^7z\, d^7z'
 G_+ (z,z')G_-(z,z')G_0(z,z')\,,
\label{GB1}
\eea
where $g^2$ is the gauge coupling constant and $G_0(z,z')$ is the
gauge superfield propagator,
\be
G_0(z,z')=
\frac1{\square}\delta^7(z-z')
=i\int_0^\infty \frac{ds}{(4\pi i
s)^{3/2}}e^{\frac{i\xi^2}{4s}}\zeta^2 \bar\zeta^2\,.
\ee
Using this propagator and the heat kernels
(\ref{K_+})--(\ref{K_-+2}), the two-loop contributions (\ref{GA1}) and (\ref{GB1}) to the effective
action can be recast as
\bea
\Gamma_{\rm A}&=&2ig^2 \int d^7z \, d^3\xi \int_0^\infty \frac{ds\, dt\,
du}{(4i\pi u)^{3/2}}
e^{i(s+t)m^2} e^{\frac{i\xi^2 }{4u}}
 K_{+-}(z,z'|s)K_{+-}(z',z|t)\big|\,,
\label{C4}
 \\
\Gamma_{\rm B}&=&2ig^2 m^2 \int d^7z \, d^3\xi \int_0^\infty
\frac{ds\, dt\, du}{(4i\pi u)^{3/2}}
e^{i(s+t)m^2} e^{\frac{i\xi^2 }{4u}}
K_+(z,z'|s)K_-(z',z|t)\big|\,.
\label{C5}
\eea
Consider first the details of computations of (\ref{C4}).

For studying the low-energy effective action up to the
four-derivative order, it is sufficient to consider the heat
kernel $K_{+-}$ in the approximation (\ref{Kapprox}),
\be
\Gamma_{\rm A}\approx\frac{2ig^2}{(4i\pi)^{9/2}} \int d^7z \, d^3\xi \int_0^\infty \frac{ds\, dt\,
du}{(st u)^{3/2}}
e^{i(s+t)(m^2+G^2)} e^{\frac{i\xi^2 }{4}(\frac1s+\frac1t+\frac1u)}
 e^{X(\xi^m,s)+X(-\xi^m,t)}\,.
\label{C6}
\ee
Using the explicit form of the function $X(\xi^m,s)$ in (\ref{X}),
we expand $e^{X(\xi^m,s)+X(-\xi^m,t)}$ in a series up to the first
order in $N_{\alpha\beta}$,
\bea
e^{X(\xi^m,s)+X(-\xi^m,t)}&=&1+i(s^2+t^2)GW^\alpha\bar W_\alpha
+\frac i3 (s^3 + t^3)GW^\alpha N_{\alpha\beta} \bar W^\beta
\nn\\&&
-\frac{s-t}{2}\xi_m \gamma^m_{\alpha\beta} W^\alpha\bar W^\beta
+\frac12(s^2-t^2) \xi_m(\gamma^m N)W^\alpha \bar W_\alpha
\nn\\&&
+\frac 32(s^2-t^2)\xi_m \gamma^m_{\gamma(\alpha}N^\gamma_{\beta)}
 W^\alpha \bar W^\beta
-\frac{7i}{24}(s^3 + t^3) W^2\bar W^2
\nn\\&&
+\frac14 G^2(s^2+t^2)^2 W^2 \bar W^2
-\frac{(s-t)^2}{16}\xi^m \xi_m W^2 \bar W^2\,.
\label{ex}
\eea

Note that some of the terms in (\ref{ex}) give no contributions to
(\ref{C6}). Indeed, the term with $GW^\alpha\bar W_\alpha$ in the r.h.s.\ of (\ref{ex})
gives vanishing contribution for considered gauge superfield
background because of (\ref{f-zero}). The terms in (\ref{ex})
linear with respect to $\xi_m$ also give vanishing contribution after
integration over $d^3\xi$ because of (\ref{int-zero}). For the remaining
terms in (\ref{ex}) we have
\bea
\Gamma_{\rm A}&\approx&\frac{2ig^2}{(4i\pi)^{9/2}} \int d^7z \, d^3\xi \int_0^\infty \frac{ds\, dt\,
du}{(st u)^{3/2}}
e^{i(s+t)(m^2+G^2)} e^{\frac{i\xi^2 }{4}(\frac1s+\frac1t+\frac1u)}
\label{C8}\\&\times&
\bigg\{
1+\frac i3(s^3+t^3)GW^\alpha N_{\alpha\beta}\bar W^\beta
+\frac{W^2\bar W^2}4\Big[
G^2(s^2+t^2)^2
-\frac{7i}{6}(s^3+t^3)
-\frac{(s-t)^2}{4}\xi^2 \Big]
\bigg\}.
\nn
\eea

The integration over $d^3\xi$ is done using (\ref{d3rho}) and
\be
\int d^3\xi\,\xi^2 e^{\frac{i}{4}a\xi^2}
=-\frac{3}{2\pi}\left(\frac{4i\pi}{a} \right)^{5/2}\,.
\ee
Then the expression (\ref{C8}) reads
\bea
\Gamma_{\rm A}&\approx&\frac{g^2}{32 \pi^3} \int d^7z \int_0^\infty \frac{ds\, dt\,
du}{(st +s u+tu)^{3/2}}
e^{i(s+t)(m^2+G^2)}
\nn\\&\times&
\bigg\{
1+\frac i3(s^3+t^3)GW^\alpha N_{\alpha\beta}\bar W^\beta
+\frac{W^2\bar W^2}4\Big[
G^2(s^2+t^2)^2\nn\\&&
-\frac{7i}{6}(s^3+t^3)
-\frac{3i}{2}\frac{(s-t)^2 stu}{st+su+tu} \Big]
\bigg\}.
\eea
After integration over $du$ we find
\bea
\Gamma_{\rm A}&\approx&\frac{g^2}{16 \pi^3} \int d^7z \int_0^\infty \frac{ds\, dt}{
\sqrt{s t}(s+t)}
e^{i(s+t)(m^2+G^2)}
\\&\times&
\bigg\{
1+\frac i3(s^3+t^3)GW^\alpha N_{\alpha\beta}\bar W^\beta
+\frac{W^2\bar W^2}4\Big[
G^2(s^2+t^2)^2
-\frac{7i}{6}(s^3+t^3)
-i\frac{(s-t)^2 st}{s+t} \Big]
\bigg\}.\nn
\eea
The remaining integrations over $s$ and $t$ can be done with the
use of the following formulas
\bea
\int_0^\infty \frac{ds\, dt}{\sqrt{st}(s+t)}e^{i(s+t)(G^2+m^2)}
&=&-\pi \ln(G^2+m^2)\,,\\
\int_0^\infty \frac{ds\, dt}{\sqrt{st}(s+t)}(s^3+t^3) e^{i(s+t)(G^2+m^2)}
&=&-\frac{5i\pi}{4(G^2+m^2)^3}\,,\\
\int_0^\infty \frac{ds\, dt}{\sqrt{st}(s+t)}(s^2+t^2)^2 e^{i(s+t)(G^2+m^2)}
&=&\frac{57\pi}{16(G^2+m^2)^4}\,,\\
\int_0^\infty\frac{ds\,dt}{(s+t)^2}\sqrt{st}(s-t)^2 e^{i(s+t)(G^2+m^2)}
&=&-\frac{i\pi}{16(G^2+m^2)^3}\,.
\eea
As a result, we get
\bea
\Gamma_{\rm A}&\approx&\frac{g^2}{16\pi^2} \int d^7z
\bigg[
-\ln(G^2+m^2)+\frac{5}{12}\frac{N_{\alpha\beta}W^\alpha \bar W^\beta}{(G^2+m^2)^3}
\nn\\&&
+\frac{W^2 \bar W^2}{96}\left(
\frac{49 G^2}{(G^2+m^2)^4}
-\frac{73}{2}\frac{m^2}{(G^2+m^2)^4}
\right)
\bigg].
\label{C16}
\eea

For computing the part of the effective action $\Gamma_{\rm B}$ up
to the four-derivative order it is sufficient to approximate the
heat kernel $K_+$ in (\ref{K+limit}) as
\be
K_+(z,z'|s)\big| \approx \frac1{(4i\pi s)^{3/2}}
s^2 W^2 e^{is G^2}e^{\frac{i\xi^2}{4s}}\,.
\label{C17}
\ee
Substituting (\ref{C17}) into (\ref{GB1}) and computing the
integrals over $d^3\xi$ and $du$ with the help of (\ref{d3rho})
one finds
\be
\Gamma_{\rm B}\approx \frac{g^2m^2}{16\pi^3}
\int d^7z\,W^2 \bar W^2\int_0^\infty
\frac{ds\,dt}{s+t}(st)^{3/2}e^{i(s+t)(G^2+m^2)}\,.
\ee
The integral over the remaining parameters reads
\be
\int_0^\infty \frac{ds\,dt}{s+t}(st)^{3/2}e^{i(s+t)(G^2+m^2)} =
\frac{9\pi}{64}\frac{1}{(G^2+m^2)^4}\,.
\ee
As a result,
\be
\Gamma_{\rm B}\approx \frac{9g^2 m^2}{1024\pi^2}\int d^7z\frac{W^2 \bar
W^2}{(G^2+m^2)^4}\,.
\label{C20}
\ee

The four-derivative two-loop effective action is given by the sum
of (\ref{C16}) and (\ref{C20}). It can be represented in the form
(\ref{eff1}) with the functions $f_i^{(2)}$ given by
\bea
f_1^{(2)} &=&-\frac{g^2}{16\pi^2}\ln(G^2+m^2)\,,\\
f_2^{(2)} &=&\frac{5 g^2}{192\pi^2}\frac{G}{(G^2+m^2)^3} \,,\\
f_3^{(2)} &=&\frac{g^2}{\pi^2}\frac{98G^2-73 m^2}{3072(G^2+m^2)^4}\,.
\eea
In sect.\ \ref{sect2.6} we denote these functions as $\tilde f_i^{(2)}$
to distinguish them form the similar functions in the $\cN=2$
Chern-Simons electrodynamics.

\end{document}